# The spectrum of $N_2$ from 4500 to 15700 cm$^{-1}$ revisited with PGOPHER

Colin M. Western[1], Luke Carter-Blatchford[1]
Patrick Crozet[2], Amanda J. Ross[2], Jérôme Morville[2]
and Dennis W. Tokaryk[3]

1  School of Chemistry, University of Bristol, BS2 8NB United Kingdom
2  Université de Lyon, Université Claude Bernard Lyon 1, CNRS, Institut Lumière Matière, F-69622, Villeurbanne, France
3 Department of Physics and Centre for Laser, Atomic and Molecular Sciences, University of New Brunswick, Fredericton, NB, Canada

**Abstract**

Using a reference molecular atlas to ensure self-consistency of wavelength calibration is widespread practice. Boesch & Reiners (Astronomy & Astrophysics 582 A43 (2015)) reported a line list from a discharge of molecular nitrogen from 4500 to 11000 cm$^{-1}$ for this purpose. With a hollow-cathode discharge source, we have extended the experimental spectrum up to 15700 cm$^{-1}$, to include the range of Ti:sapphire lasers, since the density of $N_2$ lines is greater than atomic atlases in common use. Recognizing that experimental conditions can vary, we have also analysed the spectra (comprising several B $^3\Pi_g$ – A $^3\Sigma_u^+$, B' $^3\Sigma_u^-$ – B $^3\Pi_g$, and W $^3\Delta_u$ – B $^3\Pi_g$ $N_2$ bands) with standard Hamiltonians, so that any part of the discharge spectrum 4500 – 15700 cm$^{-1}$ can be simulated. Parameters are given to do this with the spectral simulation and analysis package PGOPHER. (C. Western, J. Quant. Spectrosc. Rad. Transf., <u>186</u>, 221 (2016)). The analysis also included high-level *ab initio* calculations of potential energy curves, transition moments and spin-orbit coupling constants and these were used in preparing the model, extending the potential range of applicability. The results are available in a variety of formats to suit possible applications, including the experimental spectrum in ASCII, a detailed line list with positions and Einstein A coefficients, and a PGOPHER input file to synthesize the spectrum at selectable temperature and resolution.

## 1. Introduction

The first positive system of the $N_2$ electronic transition appears readily in most discharges containing air or nitrogen. It occurs in emission spectra from the Earth's upper atmosphere, associated with phenomena including aurorae [1], sprites [2, 3], and falling meteoroids, such as Leonid meteor showers [4]. The transitions all involve the B $^3\Pi_g$ state, with the strongest transitions to the A $^3\Sigma_u^+$ state and weaker contributions from transitions involving the B' $^3\Sigma_u^-$ and W $^3\Delta_u$ states. These electronic transitions between excited states have been extensively studied in the past, as have many other band systems of neutral and ionic forms of $N_2$, but they do not figure in standard data bases for atmospheric absorption, such as HITRAN[5]. The A $^3\Sigma_u^+$ state of $N_2$ is metastable, making it possible to detect B-A transitions in absorption in the laboratory, as well as in emission. The Lyon group became motivated to re-record the spectrum in the near-IR because they recognized contributions from the B-A system of $^{14}N_2$ around 13000 cm$^{-1}$ in an intracavity absorption experiment [6], where traces of air had leaked into a vacuum discharge cell. Although $N_2$ bands were of no spectroscopic interest in this context, they clearly had potential to provide independent calibration markers. Boesch and Reiners [7] recently pointed out that a (microwave) nitrogen discharge produces an extensive spectrum with many thousands of lines, and that spectral line lists from $N_2$ could provide useful wavelength calibration for high-resolution astrophysical spectrographs. To this end, they recorded the spectrum on a Fourier transform spectrometer, and reported transition wavenumbers and intensities, mostly without assignments, for lines in the 4500-11000 cm$^{-1}$ region. Their work stopped short of the Ti:sapphire laser wavelengths, where electronic transitions provide signatures of first row transition metal hydrides and oxides in spectra of cool stars[8]. We have extended their $N_2$ discharge 'atlas', recording emission from a hollow cathode discharge up to 15700 cm$^{-1}$ by Fourier transform spectrometry, so that the spectrum (in electronic format) could be matched to measurements from a different environment. We supply the raw spectrum, corrected for instrumental intensity response, as an ascii data file and as a simple peak list. However, because the conditions of production of $N_2$ will vary according to situation (type of discharge, pressure, temperature), comparisons will sometimes be more meaningful using a reference spectrum generated from reliable spectral parameters. The second objective of this work is therefore to present a global analysis of the transitions making up the $N_2$ atlases (taking all the unassigned lines of ref. [7], plus the new near-IR data), giving parameters from fits to what are now accepted as standard Hamiltonians for $^3\Sigma$ [9], $^3\Pi$ [10] and $^3\Delta$ [11] states, with notation following IUPAC recommendations [12]. The line assignments are unchanged from the detailed investigations on triplet transitions in $^{14}N_2$ carried out notably by groups in France in the 1970's and 1980's [13-19], but the parameters derived here have higher precision in many cases. They are otherwise essentially identical to the earlier work once the definitions of the parameters are taken into account, so we do not discuss the spectroscopic parameters in detail. We do however examine $A_D$ and γ as our increased precision allows these to be determined separately; in most cases these are completely correlated[20]. As well as conventional tables of constants, we supply the information as input for PGOPHER, a widely used program for simulating and fitting rotational, vibrational and electronic spectra [21-23].

## 2. Experiment

To complement the data published in ref. [7], we recorded $N_2$ emission from 9000 to 15700 cm$^{-1}$ at Doppler-limited resolution, on a Fourier transform spectrometer. The discharge hollow-cathode head was adapted from a sputter source [24], and consists of a cylindrical hole bored through a water-cooled copper body. Dry $N_2$ enters this system via a mass flow regulator (40 standard cm$^3$/min) through this aperture. The pressure inside the vacuum chamber was maintained around 1 mbar. The lower edge of the system has two chromium cathode plates, fixed either side of a 1.2 mm slit centred on the gas duct. The anode is a copper loop, held in place about 5 mm below the cathode. The entire source is mounted on a primary vacuum chamber, with viewing ports to send light to the spectrometer via a steering mirror and focusing lens. The discharge ran at 150 mA, 600 V for about 2½ hours to record a rotationally-resolved spectrum at an instrumental resolution of 0.023 cm$^{-1}$, averaging 80 scans. The spectrum was recorded using a Si-avalanche photodiode, which has a sharp peak in response around 960 nm. Raw intensities were therefore corrected using an instrumental intensity response function (deduced from the spectrum of a standard 100 W halogen bulb). Typical full-width half-maximum linewidths were 0.04-0.05 cm$^{-1}$, and we estimate peak position uncertainties ~ 0.005 cm$^{-1}$ for unblended lines of $N_2$. As noted by Reiners and co-workers[7] systematic calibration errors in the wavenumber scale can arise if the apertures of the FT instrument are not correctly set, or if the optical injection is imperfect. Below 11000 cm$^{-1}$, we compared peaks in our spectrum with the carefully recalibrated data from Boesch & Reiners. At higher wavenumbers, we compared peaks of atomic Cr I, Cu I and N I lines with reference wavenumbers taken from the NIST atomic data base [25], and a selection of peaks from spectra reported in [14]. A linear correction was applied to the raw wavenumber scale of our FT spectrum to minimize the differences (scattered between −0.008 and +0.021 cm$^{-1}$). Most of the N I peaks measured on our spectrum were at slightly higher wavenumber than given in NIST tables (N I lines are quoted to 2 decimal places), whilst the N$_2$ lines around 15000 cm$^{-1}$ were generally lower than those from the various spectra recorded at Laboratoire Aimé Cotton [14]. A final adjustment to the wavenumber scale of +0.001 cm$^{-1}$ (at 9000 cm$^{-1}$) to -0.004 cm$^{-1}$ (at 15500 cm$^{-1}$) was made to produce a self-consistent overall frequency scale for all the data, by ensuring that a selection of the stronger peaks measured above 11000 cm$^{-1}$ matched wavenumbers calculated from constants derived from an initial fit to the Boesch and Reiners data alone. An overview of our spectrum is shown in Figure 1.

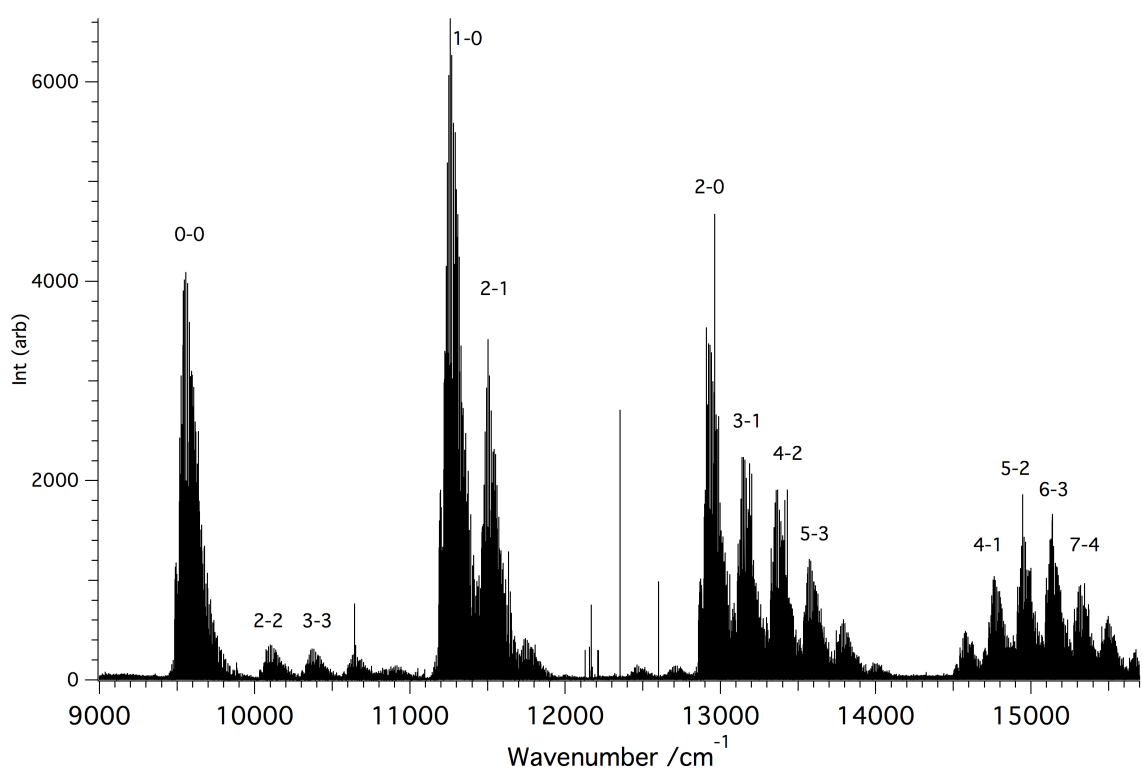

Figure 1. Overview of the Fourier transform emission spectrum of the first positive system of N$_2$, identifying the strongest bands in the B $^3\Pi_g$ – A $^3\Sigma_u^+$ system. Some N I atomic lines also appear. Intensities have been corrected for instrumental response.

### 3. Ab initio calculations

To assist in the simulation of the spectra, *ab initio* calculations for the four electronic states of interest were performed using MOLPRO 2010.1 [26, 27]. Note that the calculations were performed in D$_{2h}$ symmetry, implying that the two components of the B$^3\Pi_g$ and W$^3\Delta_u$ states have different symmetries, and both are calculated independently. Initially multi-configuration self-consistent field (MCSCF) calculations [28, 29] were performed, state-averaged over the six component states, including the two components for the $B^3\Pi_g$ and $W^3\Delta_u$ states. These were followed by multi-reference configuration interaction (MRCI) calculations [30, 31] on each of the four symmetries required based on the MCSCF calculations. The final energies used include the relaxed Davidson correction [27]. Transition dipole moments and spin-orbit matrix elements were also calculated, the later using the Breit-Pauli Hamiltonian. To investigate convergence with respect to basis set size and active space, a series of calculations were performed at a bond length, $r$, of 1.3 Å, close to the average equilibrium geometries of the four states; these are tabulated in Table S1. It is clear from the table that calculations are converged with respect to basis set and active space, with energies changing by $< 40$ cm$^{-1}$ and transition dipole moments by $< 2\%$. Calculations were thus performed using the default active space with an AV6Z basis at 72 points over the range of 0.8 – 3 Å, with the outer limit determined by convergence problems. The calculated potential energy curves, transition moments and spin-orbit matrix elements are given in tables S2-S4 of the supplementary material. RKR curves are available for these states from Roux and Michaud [19, 32], and the calculated potential energy curves are close to these, though there are some systematic differences, mainly with the RKR curves consistently shifted to slightly shorter bond length.

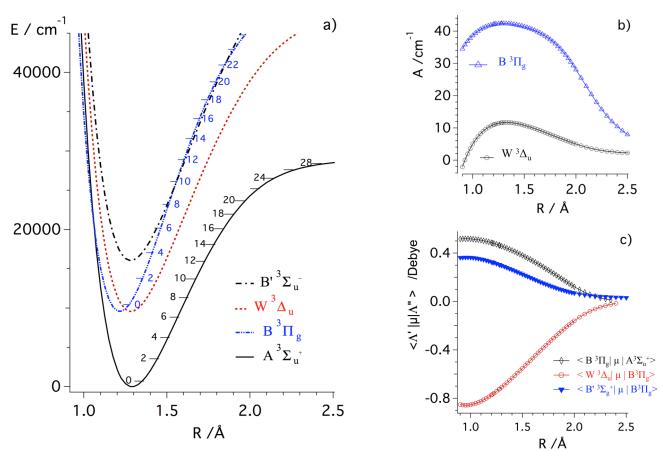

Figure 2. *Ab initio* potential energy curves for the four electronic states of N$_2$, calculated as described in the text (plot a)), together with spin-orbit functions b), and transition dipole moments, c), as a function of internuclear distance, R.

The same calculations gave spin orbit matrix elements of ~75 cm$^{-1}$ between the $A^3\Sigma^+_u$ and $B'^3\Sigma^-_u$ states, but these were ignored as they would only give an (approximately) *r*- independent shift of ~0.35 cm$^{-1}$. They will also give an equal and opposite contribution of ~0.18 cm$^{-1}$ to λ for the two states, but this is clearly not a major contribution.

The potential energy curves were fitted to an extended Morse oscillator (EMO) function [33, 34]:

$$V_{\min} + D_e\left(1 - \exp(-\beta_{\mathrm{EMO}}(r)(r - r_e))\right)^2 \quad (1)$$

where $\beta_{\mathrm{EMO}}(r)$ is expressed as a polynomial in the Šurkus reduced variable, $y(r; r_{\mathrm{ref}}, q)$:

$$\beta_{\mathrm{EMO}}(r) = \sum_{i=0} a_i y(r; r_{\mathrm{ref}}, q)^i \quad \text{where} \quad y(r; r_{\mathrm{ref}}, q) = \frac{r^q - r_{\mathrm{ref}}^q}{r^q + r_{\mathrm{ref}}^q} \quad (2)$$

Note that the reference distance, $r_{\mathrm{ref}}$, in the Šurkus reduced variable is chosen for best fit, and is typically longer than $r_e$. The number of terms in the $\beta_{\mathrm{EMO}}(r)$ polynomial were chosen to give an overall residual < 2 cm$^{-1}$. A weighted fit was used, with points above the dissociation limit being given progressively smaller weights. For the B$^3\Pi_g$ state the *ab initio* PE curve suggested a small maximum around 2.4 Å and the points around this region were also given a smaller weight. Details of the fits are given in table S2. Similar fits for the transition moments and spin-orbit coupling matrix elements are given in tables S3 and S4, though spline interpolation of the numerical values was used for the calculations below.

## 4. Data analysis and fitting with PGOPHER

The assignment and analysis of the N$_2$ spectra was straightforward, given the availability of previous analyses [13, 14]. For fitting, the Fourier transform (FT) emission spectrum described above was combined with the line list compilation given as supplementary data by Boesch and Reiners [7]. Their line list includes frequency and intensity pairs for each line, together with error estimates for both. Some lines are flagged as giving poor results in the line profile fitting process used to generate the line list, and these were discarded, but the remaining frequency and error estimates were used directly in our fits. The estimated errors in line positions were as low as 2×10$^{-6}$ cm$^{-1}$ for strong, isolated lines with 3×10$^{-5}$ cm$^{-1}$ typical for the stronger lines rising to 0.005 cm$^{-1}$ for the weakest lines. An equivalent peak list was generated from the FT spectrum, extracting error estimates with algorithms included in the PGOPHER program. The errors in peak positions based on the quality of fits of individual peaks to Gaussians were ~2×10$^{-4}$ cm$^{-1}$ for strong lines rising to 0.005 cm$^{-1}$ for the weakest lines. Peaks with large uncertainties or with widths significantly larger than the average were discarded. Given the calibration procedure described in the experimental section we are confident that the absolute accuracy is better than 0.01 cm$^{-1}$ even at the highest wavenumbers recorded. The fits below demonstrate the relative accuracy is much better than this, though not quite as good as the estimates based on fitting individual line profiles.

As discussed below, an important part of the assignment process involves comparison of calculated and observed intensities. The fluorescence intensities were calculated from the Einstein A coefficients for the transition (derived from the *ab initio* dipole moment and potential energy curves for the purpose of initial assignment), weighted by the Boltzmann factor for the upper state. The calculated area is then proportional to the number of photons emitted per second on the transition, requiring a correction factor proportional to frequency for the experimental areas, as these are proportional to emitted power. The Boesch and Reiners data gives peak width as well as intensity at the line centre, so the product of these two is taken to give the peak area. An additional correction is required, as their intensity calibration is done against a black body curve expressed as energy per wavelength, and the areas must by divided by frequency cubed to give a value proportional to the rate of emission of photons.

The automatic assignment feature of PGOPHER, described in ref. [22] was then used to assign these line lists. Given the previous analyses, an extensive search is not required and in practice most assignment was done by simply taking the line nearest to each calculated line, excluding obvious mis-assignments, fitting and repeating the process a couple of times with re-calculated positions of unassigned lines. Given the large number of lines assigned identifying mis-assignments requires computational assistance. The obvious mechanism of excluding lines with residuals above a certain threshold is rather crude, as it gives no assurance that lines are being excluded for good reasons. We have taken an alternative approach, based on implementing special intensity residuals plots in the PGOPHER program. This new feature offers various (user selectable) plots based on observed and calculated intensity, and allows transitions outside the expected intensity range to be easily discarded. A key feature of the method is that only transitions with unexpectedly high intensity are discarded, while transitions with observed intensity unexpectedly low give cause for concern. Observed intensity too high simply corresponds to overlapping transitions and is frequently encountered because of the line density. If the fit is correct, all transitions above some intensity threshold should appear in the spectrum, and any absence suggests a mis-assignment or problem with the model. Figure 3 shows typical plots, in this case for the origin band of the $B\ ^3\Pi_g - A\ ^3\Sigma_u^+$ transition using data from [7]. Both plots show the log of the ratio of intensity of the experimental spectrum at the line centre to the calculated line strength, *S*, against the upper state energy. Note that we use the measured line position as line centre for assigned features, and take the associated intensity at that point on the spectrum, rather than attempt to use measured peak heights. Calculated line centres, and the intensity on the experimental spectrum, are used for unassigned (or missing) transitions. The intensity plots can then draw attention to missing, or unexpectedly weak, lines in the spectrum. The plots shown in Fig. 3 use the calculated line strength (which excludes the Boltzmann factor), rather than the overall calculated intensity, to give a ratio proportional to population, and so under ideal conditions a linear Boltzmann plot is expected. (The calculated line intensity can also be used for these plots, in which case a horizontal line is expected, though the relatively large scatter in measured intensities made the plots shown more convenient in the current study.)

Figure 3(b) includes only the 437 transitions included in the final fit, and is close to the expected linear form, though the plot shows a clear curvature arising from a non-Boltzmann population distribution discussed in more detail below. Figure 3(a) includes all 978 calculated transitions with intensities more than 0.3% of the strongest, and shows many transitions stronger than calculated but, crucially, none weaker than calculated, implying all the discrepancies can be understood in terms of overlapping transitions.

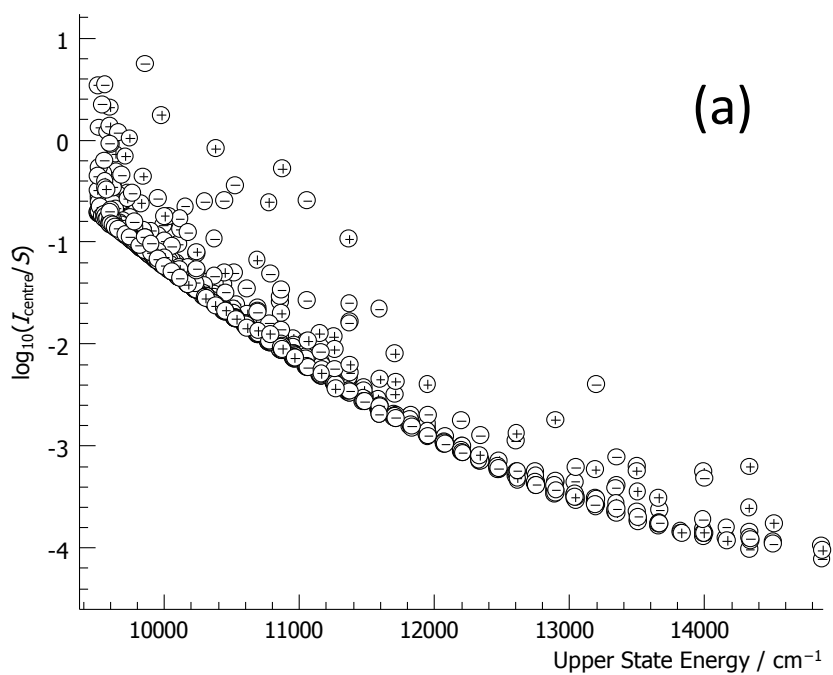

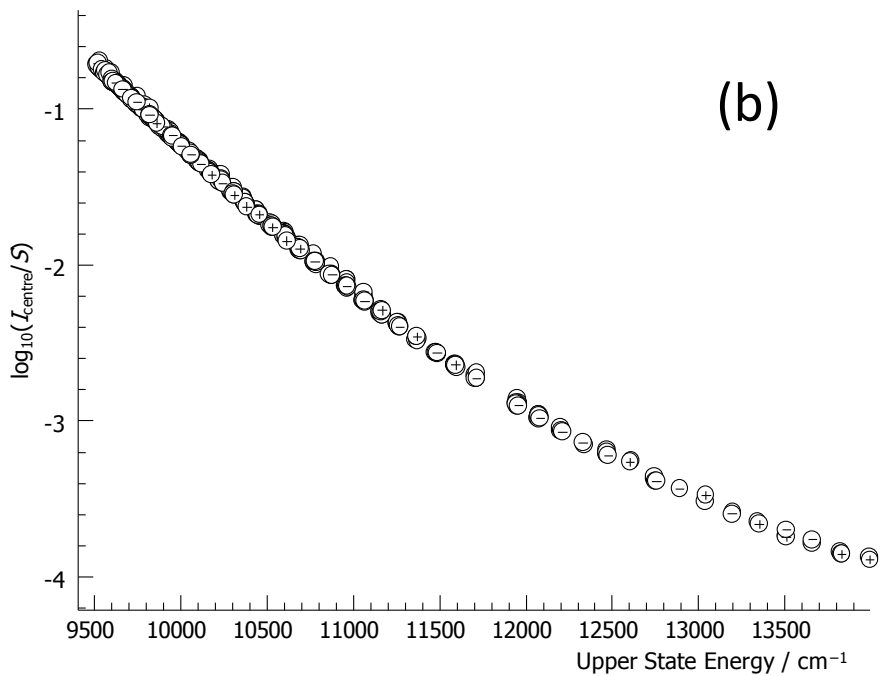

Figure 3. Plots of the log of the ratio of experimental intensity at the line centre to calculated line strength against upper state energy for the origin band of the $B\ ^3\Pi_g - A\ ^3\Sigma_u^+$ transition using data from [7]. (a) includes all 978 transitions, and (b) only those 437 included in the final fit. Note that, as discussed in the text, there is a sharp lower limit at each energy. + and – symbols indicate e and f parity in the upper state.

The transitions were selected so that most of the observed/calculated intensity ratios were within ±10% of our final intensity model, with more latitude (up to 30%) given to the weak, high-$J$ transitions. This pruned the line list quite significantly, and, for the band shown, only 45% of the transitions were retained. This may seem to be discarding much relevant information, but given the large number of lines assigned there are still sufficient observations for a good determination of the (up to 20) parameters for each band, particularly as the coverage of $J$, $\Omega$ and parity was even. It is important to be aware of this, as the discarded lines included some of the strongest, where a blend led to a poor fit to a Gaussian function or an anomalously high intensity, and such lines would typically be included in standard fits, though usually with reduced weights. Even after the selection by intensity, a few lines with large residuals remained (and were discarded) as, with our data quality, even a blend with a line of intensity 10% of the stronger line (and thus satisfying the intensity criterion) experiences a sufficiently large shift to be visible in the fit, and thus merits discarding. The intensity pruning was iterative, as a good intensity model is required for the final pruning, which was in turn generated as part of the modelling process; the intensities are discussed below. Combining line lists from our work and Boesch and Reiners, assignments were made for v = 0 – 9 for the $B\ ^3\Pi_g$ state, v = 0 – 8 for the $A\ ^3\Sigma_u^+$ state, v = 0 – 7 for the $B'\ ^3\Sigma_u^-$ state and v = 0 – 12 for the $W\ ^3\Delta_u$ state. While the measurements here have
9

significantly higher precision than the earlier work from Roux et al [14, 18, 19, 32], the coverage in v is smaller, reflecting different discharge conditions. Overall most of the transitions in the spectra are modelled, with some atomic lines from Cr (the cathode) and N are also present (the latter easily recognizable from their hyperfine-broadened profiles), and some transitions from the A – X transition in $N_2^+$.

The Hamiltonian used to model this data is as standard as possible [12, 21], and includes rotational, spin-orbit, spin-spin, spin-rotation and Λ-doubling terms, with centrifugal distortion included as necessary to fit the data:

$$H = H_{rot} + H_{SO} + H_{SS} + H_{SR} + H_{LD} \tag{3}$$

$$H_{rot} = B\hat{\mathbf{N}}^2 - D\hat{\mathbf{N}}^4 + H\hat{\mathbf{N}}^6 \tag{4}$$

$$H_{SO} = A\hat{L}_z\hat{S}_z + \frac{A_D}{2}\left[\hat{\mathbf{N}}^2, \hat{L}_z\hat{S}_z\right]_+ \tag{5}$$

$$H_{SS} = \tfrac{2}{3}\lambda\left(3\hat{S}_z^2 - \hat{\mathbf{S}}^2\right) + \tfrac{1}{2}\lambda_D\left[\tfrac{2}{3}\left(3\hat{S}_z^2 - \hat{\mathbf{S}}^2\right), \hat{\mathbf{N}}^2\right]_+ + \tfrac{1}{2}\lambda_H\left[\tfrac{2}{3}\left(3\hat{S}_z^2 - \hat{\mathbf{S}}^2\right), \hat{\mathbf{N}}^4\right]_+ \tag{6}$$

$$H_{SR} = \gamma\hat{\mathbf{N}}\cdot\hat{\mathbf{S}} + \tfrac{1}{2}\gamma_D\left[\hat{\mathbf{N}}\cdot\hat{\mathbf{S}}, \hat{\mathbf{N}}^2\right]_+ \tag{7}$$

$$\begin{aligned}H_{LD} =\ & \tfrac{1}{2}o\left(\hat{S}_+^2 e^{-2i\varphi} + \hat{S}_-^2 e^{+2i\varphi}\right) + \tfrac{1}{4}o_D\left[\hat{S}_+^2 e^{-2i\varphi} + \hat{S}_-^2 e^{+2i\varphi}, \hat{\mathbf{N}}^2\right]_+ \\ & - \tfrac{1}{2}p\left(\hat{N}_+\hat{S}_+ e^{-2i\varphi} + \hat{N}_-\hat{S}_- e^{-2i\varphi}\right) - \tfrac{1}{4}p_D\left[\hat{N}_+\hat{S}_+ e^{-2i\varphi} + \hat{N}_-\hat{S}_- e^{-2i\varphi}, \hat{\mathbf{N}}^2\right]_+ \\ & + \tfrac{1}{2}q\left(\hat{N}_+^2 e^{-2i\varphi} + \hat{N}_-^2 e^{-2i\varphi}\right) + \tfrac{1}{4}q_D\left[\hat{N}_+^2 e^{-2i\varphi} + \hat{N}_-^2 e^{-2i\varphi}, \hat{\mathbf{N}}^2\right]_+\end{aligned} \tag{8}$$

Initially, a "traditional" fit to the data was performed, fitting each of the above parameters as required for each state. This gave good results, with 17967 transitions fitted to 298 parameters with an average unweighted error of 0.0023 cm$^{-1}$ and no evidence of perturbations. See tables 1-4 for the parameters. Two points are worthy of comment. Firstly, the fit was a weighted fit, with weights for the observed frequencies derived from the fits of individual line profiles. The fit to the lines from reference [7] gave a weighted (dimensionless) average error of 7.1 and an unweighted average error of 0.0019 cm$^{-1}$, whereas the fit to our data gave a weighted average error of 3.3 and an unweighted average error of 0.0030 cm$^{-1}$. This implies our data is slightly less precise, which can be understood in terms of the greater Doppler width at higher frequency, and a slightly poorer signal to noise ratio. However, the weighted errors imply the uncertainty estimated (in both cases) from the fits to line profiles is too small, by a factor of 3 to 7. To balance the two data sets, the estimated errors from reference [7] were scaled up by a factor of 3, and the overall weighted error was then 2.7. The overall unweighted error was 0.0023 cm$^{-1}$ though this is a slightly misleading figure given the wide range of weights used. For strong unblended lines the typical error was $2\times10^{-4}$ cm$^{-1}$ for the data from reference [7] and $6\times10^{-4}$ cm$^{-1}$ above 11000 cm$^{-1}$.

Table 1 Constants (/cm$^{-1}$) for the A$^3\Sigma^+_u$ state of N$_2$

| v | Origin | | λ | | B | | D×10$^6$ | | γ×10$^3$ | | λ$_D$×10$^6$ | H×10$^{12}$ |
|---|---|---|---|---|---|---|---|---|---|---|---|---|
| | This Work[a] | Previous[b] | This Work | Previous[b,c] | This Work | Previous[b] | This Work | Previous[b] | This Work | Previous[b] | | |
| 0 | 0[d] | 0 | -1.328301(14) | -1.32855(42) | 1.445765671(219) | 1.445774(6) | 5.794165(384) | 5.804(2) | -2.71836(55) | -2.4(2) | 3.690(59) | -2.259(199) |
| 1 | 1432.907215(11) | 1432.9066 | -1.321342(11) | -1.32165(42) | 1.427459855(215) | 1.427440(10) | 5.834309(377) | 5.827(5) | -2.69331(51) | -2.5(2) | 3.728(53) | -2.892(197) |
| 2 | 2838.142474(15) | 2838.1400 | -1.314253(16) | -1.31430(64) | 1.409051702(218) | 1.409044(17) | 5.877523(383) | 5.87(1) | -2.66775(59) | -2.1(3) | 3.743(66) | -3.822(201) |
| 3 | 4215.637926(22) | 4215.6347 | -1.307003(20) | -1.30770(42) | 1.390528368(219) | 1.390530(8) | 5.924195(329) | 5.935(4) | -2.64054(74) | -2.3(2) | 3.908(90) | -5.000(146) |
| 4 | 5565.297562(30) | 5565.2917 | -1.299521(31) | -1.30020(67) | 1.371875068(283) | 1.371905(14) | 5.975698(524) | 5.995(7) | -2.61286(95) | -2.7(2) | 3.91(12) | -6.158(288) |
| 5 | 6886.992792(41) | 6886.9847 | -1.291810(47) | -1.29225(67) | 1.353073432(365) | 1.353090(9) | 6.033523(721) | 6.052(5) | -2.5804(13) | -2.3(4) | 4.13(17) | -6.733(411) |
| 6 | 8180.556724(70) | 8180.5487 | -1.283961(85) | -1.28400(54) | 1.334101621(646) | 1.334097(14) | 6.09710(148) | 6.115(9) | -2.5498(19) | -2.1(2) | 4.08(28) | -9.216(946) |
| 7 | 9445.77848(11) | 9445.7557 | -1.27558(13) | -1.27500(64) | 1.31493401(101) | 1.314909(17) | 6.17160(254) | 6.160(8) | -2.5191(28) | -2.2(3) | 4.18(42) | -9.62(179) |
| 8 | 10682.39534(17) | 10682.3770 | -1.26680(25) | -1.26735(75) | 1.29554387(107) | 1.295550(19) | 6.2695(14) | 6.29(1) | -2.4903(48) | -2.3(4) | 3.95(80) | 0[d] |
| 9 | | 11890.0380 | | -1.2575(15) | | 1.27597(3) | | 6.41(1) | | -2.36(5) | | |
| 10 | | 13068.4170 | | -1.2484(15) | | 1.25595(3) | | 6.51(2) | | -2.24(12) | | |
| 11 | | 14217.0100 | | -1.2390(15) | | 1.23555(5) | | 6.62(3) | | -3.03(10) | | |
| 12 | | 15335.2800 | | -1.2239(30) | | 1.21496(4) | | 6.89(2) | | -2.96(8) | | |
| 13 | | 16422.5320 | | -1.2183(15) | | 1.19370(4) | | 7.02(2) | | -3.18(4) | | |
| 14 | | 17478.0070 | | -1.2048(30) | | 1.17193(5) | | 7.25(2) | | -3.20(8) | | |
| 15 | | 18501.2760 | | -1.1904(30) | | 1.14940(7) | | 7.49(3) | | -3.60(10) | | |
| 16 | | 19490.4440 | | -1.1829(45) | | 1.12619(10) | | 7.82(4) | | -3.57(12) | | |

Footnotes

a Systematic error may be up to 0.01 cm$^{-1}$. Value for v = 0 is fixed at zero.

b Ref [14] for v=0-8, [19] for v = 9-16

c Value quoted is 3/2(ε+γ) from previous work

d Fixed

Table 2 Constants (/cm$^{-1}$) for the B$^3\Pi_g$ state of N$_2$

| v | Origin | | A | | λ | | o | | B | | D×10$^6$ | | p×10$^3$ | |
|---|---|---|---|---|---|---|---|---|---|---|---|---|---|---|
| | This Work[a] | Previous[b,c] | This Work | Previous[b] | This Work | Previous[b,d] | This Work | Previous[b,e] | This Work | Previous[b] | This Work | Previous[b] | This Work | Previous[b,f] |
| 0 | 9550.0078976(553) | 9551.6468 | 42.2388809(418) | 42.2321(14) | -0.202318(20) | -0.2025(8) | 1.1524340(49) | 1.1517(5) | 1.628768629(213) | 1.628799(10) | 5.861665(370) | 5.860(4) | 4.3438(12) | 4.32(18) |
| 1 | 11255.202011(101) | 11256.8344 | 42.1935541(763) | 42.1845(7) | -0.203904(35) | -0.2042(5) | 1.150064(10) | 1.1496(3) | 1.610550103(192) | 1.610583(6) | 5.895879(217) | 5.897(2) | 4.3148(22) | 4.31(17) |
| 2 | 12931.534005(155) | 12933.1455 | 42.143467(118) | 42.135(1) | -0.205684(54) | -0.2057(6) | 1.147688(18) | 1.1471(3) | 1.592221358(234) | 1.592235(9) | 5.933394(277) | 5.930(5) | 4.2982(34) | 4.29(17) |
| 3 | 14578.944926(276) | 14580.5378 | 42.088527(211) | 42.078(2) | -0.207413(96) | -0.2083(8) | 1.145243(37) | 1.1448(4) | 1.573778783(314) | 1.57378(1) | 5.973763(393) | 5.958(5) | 4.2829(60) | 4.25(17) |
| 4 | 16197.371505(335) | 16198.9478 | 42.029070(258) | 42.022(2) | -0.20940(13) | -0.2088(12) | 1.142761(47) | 1.1422(6) | 1.555217538(383) | 1.55525(2) | 6.01533(47) | 6.03(2) | 4.2597(73) | 4.25(16) |
| 5 | 17786.744674(499) | 17788.3048 | 41.965039(388) | 41.953(2) | -0.21071(18) | -0.2121(14) | 1.140403(71) | 1.1397(6) | 1.536532747(511) | 1.53655(2) | 6.05969(68) | 6.06(1) | 4.203(11) | 4.24(16) |
| 6 | 19346.993715(694) | 19348.5266 | 41.892098(527) | 41.882(2) | -0.21313(26) | -0.2123(14) | 1.13775(11) | 1.1372(8) | 1.517719992(816) | 1.51776(2) | 6.1081(12) | 6.11(1) | 4.211(15) | 4.22(16) |
| 7 | 20878.02608(23) | 20879.5396 | 41.81553(25) | 41.808(3) | -0.21568(25) | -0.2170(15) | 1.13466(22) | 1.1337(7) | 1.49876772(172) | 1.49880(1) | 6.1574(24) | 6.160(7) | 4.127(18) | 4.18(15) |
| 8 | 22379.74542(35) | 22381.2386 | 41.72842(37) | 41.718(3) | -0.21703(38) | -0.2215(15) | 1.13315(31) | 1.1313(7) | 1.47966848(259) | 1.47970(2) | 6.1961(38) | 6.21(1) | 4.225(24) | 4.18(15) |
| 9 | 23852.02913(187) | 23853.5076 | 41.62742(210) | 41.625(1) | -0.2289(13) | -0.220(3) | 1.12922(50) | 1.1292(8) | 1.4604439(155) | 1.46045(2) | 6.2909(330) | 6.30(2) | 4.103(80) | 4.12(15) |
| 10 | | 25296.2036 | | 41.525(3) | | -0.223(3) | | 1.1250(10) | | 1.44098(3) | | 6.31(2) | | 4.11(14) |
| 11 | | 26709.1671 | | 41.415(1) | | -0.2243(12) | | 1.1247(8) | | 1.42137(2) | | 6.39(1) | | 4.10(14) |
| 12 | | 28092.2091 | | 41.292(2) | | -0.2250(14) | | 1.1202(7) | | 1.40157(3) | | 6.49(1) | | 4.07(14) |
| 13 | | 29445.069 | | 41.166(3) | | -0.241(3) | | 1.125(2) | | 1.38158(5) | | 6.61(2) | | 4.12(17) |
| 14 | | 30767.488 | | 41.011(3) | | -0.241(9) | | 1.122(3) | | 1.36122(3) | | 6.69(2) | | 4.03(23) |
| 15 | | 32059.121 | | 40.862(4) | | -0.237(6) | | 1.120(4) | | 1.34047(5) | | 6.72(4) | | 4.62(24) |
| 16 | | 33319.556 | | 40.667(3) | | -0.245(5) | | 1.107(2) | | 1.31958(7) | | 7.02(5) | | 3.62(17) |
| 17 | | 34548.266 | | 40.461(2) | | -0.251(3) | | 1.106(2) | | 1.29831(4) | | 7.17(2) | | 3.82(17) |
| 18 | | 35744.634 | | 40.226(2) | | -0.243(3) | | 1.105(2) | | 1.27639(4) | | 7.30(2) | | 3.70(14) |
| 19 | | 36907.953 | | 39.969(4) | | -0.257(5) | | 1.099(3) | | 1.25401(6) | | 7.54(3) | | 3.70(14) |
| 20 | | 38037.353 | | 39.677(5) | | -0.259(5) | | 1.094(3) | | 1.23087(7) | | 7.79(3) | | 3.36(13) |
| 21 | | 39131.827 | | 39.332(7) | | -0.269(11) | | 1.087(4) | | 1.20708(10) | | 8.17(4) | | 3.33(17) |

Footnotes
a Systematic error may be up to 0.01 cm$^{-1}$.
b Ref [32] for v=0-12, [19] for v = 13-21
c Value quoted is Origin+B from previous work
d Value quoted is 3/2(ε−p/2) from previous work
e Value quoted is −α from previous work
f Value quoted is ½(p−2q) from previous work
g Value quoted is −2(q+D) from previous work
h Value from previous work not directly comparable
i Fixed

Table 2 continued. Constants (/cm$^{-1}$) for the B$^3\Pi_g$ state of N$_2$

| v | $q\times10^5$ | | $\gamma\times10^3$ | $A_D\times10^4$ | $\lambda_D\times10^7$ | $\gamma_D\times10^8$ | $q_D\times10^9$ | $H\times10^{12}$ |
|---|---|---|---|---|---|---|---|---|
|  | This Work | Previous[b,g] | This Work[h] | This Work[h] | This Work | This Work | This Work | This Work |
| 0 | 8.4703(50) | 8.52(10) | -3.62240(3797) | -4.21312(3174) | -6.73(72) | -3.653(93) | 1.552(64) | 1.130(194) |
| 1 | 8.268(12) | 8.26(6) | -3.85913(6945) | -4.38467(5736) | -4.4(13) | -3.46(22) | 1.14(16) | 0[i] |
| 2 | 7.975(21) | 8.05(10) | -3.7324(1058) | -4.25834(8650) | -7.9(22) | -4.29(38) | 2.21(31) | 0[i] |
| 3 | 7.758(48) | 7.90(20) | -3.4640(1895) | -4.0164(1530) | -14.3(40) | -3.35(85) | 2.18(73) | 0[i] |
| 4 | 7.694(26) | 7.60(30) | -3.5400(2306) | -4.0594(1841) | 0[i] | -2.6(11) | 0[i] | 0[i] |
| 5 | 7.402(39) | 7.35(20) | -4.7047(3460) | -4.9589(2743) | 0[i] | 0[i] | 0[i] | 0[i] |
| 6 | 7.01(17) | 7.29(30) | -3.1417(4796) | -3.7114(3747) | 0[i] | 0[i] | 0[i] | 0[i] |
| 7 | 7.05(16) | 6.94(20) | -3.765(16) | -4.21312[i] | 0[i] | 0[i] | 0[i] | 0[i] |
| 8 | 7.97(21) | 6.99(20) | -3.900(25) | -4.21312[i] | 0[i] | 0[i] | 0[i] | 0[i] |
| 9 | 8.5(13) | 6.74(30) | -4.26(14) | -4.21312[i] | 0[i] | 0[i] | 0[i] | 0[i] |
| 10 |  | 6.33(30) |  |  |  |  |  |  |
| 11 |  | 6.53(30) |  |  |  |  |  |  |
| 12 |  | 6.33(50) |  |  |  |  |  |  |
| 13 |  | 6.32(30) |  |  |  |  |  |  |
| 14 |  | 8.1(10) |  |  |  |  |  |  |
| 15 |  | 6.2(20) |  |  |  |  |  |  |
| 16 |  | 3.6(14) |  |  |  |  |  |  |
| 17 |  | 3.54(90) |  |  |  |  |  |  |
| 18 |  | 4.44(80) |  |  |  |  |  |  |
| 19 |  | 4.67(80) |  |  |  |  |  |  |
| 20 |  | 4.01(70) |  |  |  |  |  |  |
| 21 |  | 3.2(10) |  |  |  |  |  |  |

Table 3 Constants (/cm$^{-1}$) for the W$^3\Delta_u$ state of N$_2$

| v | Origin | | A | | λ | | B | | D×10$^6$ | | γ×10$^3$ | |
|---|--------|--------|---|---|---|---|---|---|---|---|---|---|
| | This Work[a] | Previous[b,c] | This Work | Previous[b] | This Work | Previous[b,d] | This Work | Previous[b] | This Work | Previous[b] | This Work | Previous[b,e] |
| 0 | 9619.60060(65) | 9625.4504(32) | 5.8174(12) | 5.8141(13) | 0.66597(61) | 0.66975(60) | 1.46180782(445) | 1.461743(17) | 5.6435(63) | 5.608(12) | -2.937(31) | -2.915(45) |
| 1 | 11101.13790(69) | 11106.9165(17) | 5.76283(90) | 5.7568(14) | 0.67042(60) | 0.67395(60) | 1.44473840(495) | 1.444687(13) | 5.6239(74) | 5.616(7) | -3.005(24) | -2.970(40) |
| 2 | 12557.80477(48) | 12563.5179(6) | 5.71082(64) | 5.7096(8) | 0.67534(33) | 0.67515(45) | 1.42768803(318) | 1.427643(10) | 5.64479(447) | 5.646(5) | -2.913(14) | -2.805(30) |
| 3 | 13989.73512(24) | 13995.3789(7) | 5.65683(39) | 5.6542(13) | 0.67848(27) | 0.68040(45) | 1.4106314(17) | 1.410595(14) | 5.6494(23) | 5.665(7) | -2.877(11) | -3.035(40) |
| 4 | 15397.04969(45) | 15402.6254(17) | 5.60023(62) | 5.6007(11) | 0.68143(37) | 0.68325(60) | 1.39358996(324) | 1.393536(17) | 5.6767(50) | 5.677(10) | -2.850(18) | -2.940(38) |
| 5 | 16779.86826(38) | 16785.3759(41) | 5.54266(71) | 5.5461(20) | 0.68476(31) | 0.68760(75) | 1.37653387(196) | 1.376489(19) | 5.6935(21) | 5.696(9) | -2.807(11) | -3.188(68) |
| 6 | 18138.28951(24) | 18143.7291(27) | 5.47991(42) | 5.4785(21) | 0.68894(22) | 0.69075(90) | 1.35947620(137) | 1.359430(29) | 5.7159(15) | 5.715(23) | -2.7953(87) | -3.273(78) |
| 7 | 19472.40489(39) | 19477.7802(16) | 5.41716(59) | 5.4357(22) | 0.69256(30) | 0.69255(75) | 1.34240175(246) | 1.34233(3) | 5.7368(31) | 5.74(2) | -2.767(13) | -2.625(75) |
| 8 | 20782.28719(36) | 20787.5820 | 5.35270(53) | 5.3427(36) | 0.69595(25) | 0.697(15) | 1.32530630(216) | 1.32522(6) | 5.7585(28) | 5.74(2) | -2.757(12) | -2.050(50) |
| 9 | 22067.99104(31) | 22073.214 | 5.28740(57) | 5.2676(40) | 0.69973(25) | 0.700(15) | 1.30818187(173) | 1.30815(6) | 5.7866(19) | 5.80(5) | -2.695(12) | -2.200(25) |
| 10 | 23329.54814(51) | 23334.696 | 5.21759(77) | 5.218(30) | 0.70315(32) | 0.702(15) | 1.29101360(263) | 1.29095(1) | 5.81297(292) | 5.8(1) | -2.678(13) | -2.300(25) |
| 11 | 24566.96640(50) | 24572.035 | 5.14572(99) | 5.138(50) | 0.70514(39) | 0.709(30) | 1.27379929(248) | 1.27384(1) | 5.8511(27) | 6.0(1) | -2.648(16) | -1.950(25) |
| 12 | 25780.22595(97) | 25785.200 | 5.0712(17) | 5.042(50) | 0.70899(93) | 0.706(45) | 1.25652651(623) | 1.25688(2) | 5.9052(85) | 5.7(2) | -2.669(39) | -2.0(5) |

Footnotes

a Systematic error may be up to 0.01 cm$^{-1}$.
b Ref [18] for v=0-6, [32] for v = 7-12
c Value quoted is Origin − 4B from previous work
d Value quoted is 3/2ε from previous work
e Value quoted is −4p from previous work

Table 4 Constants (/cm$^{-1}$) for the B'$^3\Sigma^-_u$ state of N$_2$

| v | Origin | | λ | | B | | D×10$^6$ | | γ×10$^3$ | | λ$_D$×10$^6$ | γ$_D$×10$^8$ | H×10$^{12}$ |
|---|---|---|---|---|---|---|---|---|---|---|---|---|---|
| | This Work[a] | Previous[b,c] | This Work | Previous[b,d] | This Work | Previous[b] | This Work | Previous[b] | This Work | Previous[b] | | | |
| 0 | 16096.181701(27) | 16096.1817 | 0.652871(35) | 0.65267(32) | 1.464775049(295) | 1.464777(06) | 5.564059(564) | 5.563(3) | -2.5320(15) | -2.58(15) | 7.9(12) | 1.43(31) | 1.980(304) |
| 1 | 17588.860591(26) | 17588.8604 | 0.654386(34) | 0.65366(48) | 1.448143106(296) | 1.448158(12) | 5.572463(581) | 5.584(11) | -2.5238(14) | -2.66(22) | 9.9(11) | 1.97(28) | 1.335(323) |
| 2 | 19057.597867(75) | 19057.6090 | 0.655992(89) | 0.65677(78) | 1.431538681(662) | 1.431532(20) | 5.58302(145) | 5.572(20) | -2.5112(18) | -2.51(30) | 10.5(27) | 0$^e$ | 1.956(884) |
| 3 | 20502.564765(85) | 20502.5751 | 0.657788(89) | 0.65705(67) | 1.414959202(518) | 1.414970(20) | 5.58947(61) | 5.604(20) | -2.5047(28) | -2.94(40) | 0$^e$ | 0$^e$ | 0$^e$ |
| 4 | 21923.92787(12) | 21923.9365 | 0.65917(12) | 0.66037(75) | 1.398408069(718) | 1.398422(26) | 5.59784(85) | 5.611(20) | -2.5006(34) | -2.11(6) | 0$^e$ | 0$^e$ | 0$^e$ |
| 5 | 23321.84911(27) | 23321.8586 | 0.66069(23) | 0.66325(75) | 1.38188262(166) | 1.381803(24) | 5.6055(21) | 5.512(15) | -2.4977(77) | -1.87(5) | 0$^e$ | 0$^e$ | 0$^e$ |
| 6 | 24696.48305(34) | 24696.4963 | 0.66190(30) | 0.6621(30) | 1.36538523(196) | 1.36534(8) | 5.6173(23) | 5.30(10) | -2.4917(98) | -5.26(12) | 0$^e$ | 0$^e$ | 0$^e$ |
| 7 | 26047.97536(51) | 26047.9833 | 0.66349(47) | 0.6619(15) | 1.34890931(321) | 1.34888(4) | 5.6256(41) | 5.570(80) | -2.502(15) | -3.43(6) | 0$^e$ | 0$^e$ | 0$^e$ |
| 8 | | 27376.4633 | | 0.6647(10) | | 1.33246(2) | | 5.650(20) | | -3.6(5) | | | |
| 9 | | 28682.0533 | | 0.6650(16) | | 1.31601(6) | | 5.730(60) | | -2.3(9) | | | |

Footnotes
a Systematic error may be up to 0.01 cm$^{-1}$
b Ref [32] (1988)
c Offset by 9551.6433 cm$^{-1}$
d Value quoted is 3/2(ε+γ/2) from previous work
e Fixed

Given the precision of the fits, with residuals for the strongest lines down to 1/200 of the linewidth it is worth considering the effects of the unresolved hyperfine structure. For the B $^3\Pi_g$ – A $^3\Sigma_u^+$ transition the hyperfine structure has been measured [35], and the structure of an individual rotational transition is typically spread over 0.003 to 0.015 cm$^{-1}$, depending on the branch and $J$. This is potentially of concern, given the typical Doppler linewidth of ~0.04 cm$^{-1}$, but tests with fitting a Doppler-broadened exact line profile to a simple Gaussian indicates the potential shifts are very small. Even for a transition with a large hyperfine splitting, such as $^SR_{32}(1)$ where the components are spread over 0.014 cm$^{-1}$, the shift is only $10^{-5}$ cm$^{-1}$, so this potential systematic error can be ignored.

The inclusion of both the centrifugal distortion of the spin-orbit coupling, $A_D$ and the spin-rotation, $\gamma$ in the fit for the B $^3\Pi_g$ state also deserves special mention. It is well known that these two parameters are strongly correlated (see for example [20] for a formal derivation), and normally only one of these is determined. However, the argument for this correlation is based on perturbation theory, and a numerical test based on exact calculations (given in the appendix) shows that the effects of these two parameters are not exactly identical, and our precision is just sufficient to determine both independently for selected states. For the origin band of the B $^3\Pi_g$ – A $^3\Sigma_u^+$ transition, for which the best data is available, fitting to just $A_D$ or $\gamma$ gives significantly worse fits, as shown in Table 5, with weighted residuals plotted in Figure 4. This test fit just included the data from reference [7], 437 lines in all with $J$ up to 53.

Table 5. Results of alternative fits to the origin band of the B $^3\Pi_g$ – A $^3\Sigma_u^+$ transition.

|  | Fit Both | $\gamma = 0$ | $A_D = 0$ |
|---|---|---|---|
| $\gamma \times 10^4$ / cm$^{-1}$ | -36.0795(9562) | - | 14.230(56) |
| $A_D \times 10^4$ / cm$^{-1}$ | -4.20317(7994) | -1.1895(36) | - |
| Average Weighted Error | 3.2 | 6.7 | 8.9 |

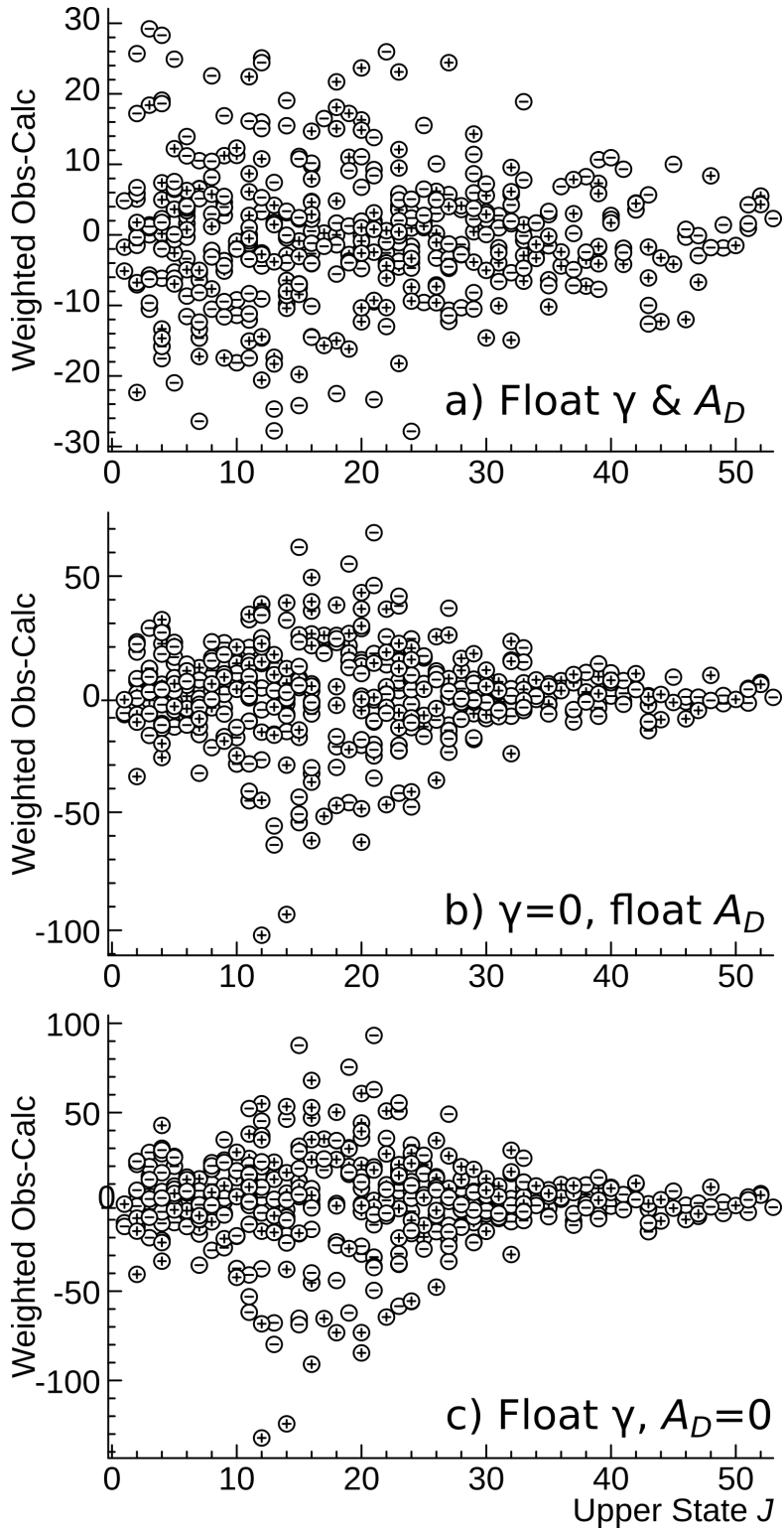

Figure 4. Plot of the weighted residuals for the three alternative fits given in table 5 to the origin band of the B $^3\Pi_g$ – A $^3\Sigma_u^+$ transition in $N_2$. The top trace corresponds to floating both $\gamma$ and $A_D$, the middle fixing $\gamma$ at zero and the bottom fixing $A_D$ to zero. Note the different vertical scales in the three plots.

There is still a strong correlation between the two parameters, as can be seen by the requirement, based on Watson's sensitivity parameter [36], for several significant figures in the standard deviation where both are floated. The previous workers on the B $^3\Pi_g$ state[14] were aware of this issue, and constrained one or the other to zero, though their precision on these constants was much less, and in many cases they could determine neither. An additional consideration is that, as discussed more generally below, $A_D$ can be calculated directly from the ab initio $A(r)$ curve, and this gives values $\sim 2\times 10^{-5}$ cm$^{-1}$, some 1/20$^{\text{th}}$ of the observed values. This implies the major contribution to $A_D$ is not centrifugal distortion, and $A_D$ is an effective parameter. Looking at the other parameters, $A_D \sim -p$, and a possible contribution to $A_D$ can easily be identified. $p$ arises from the $L$-uncoupling interaction, $\langle B\ ^3\Pi_g | B\hat{J}_+\hat{L}_- + B\hat{J}_-\hat{L}_+ | ^3\Sigma_g \rangle$, with other $^3\Sigma_g$ states, and if the same term is used to mix with other $^3\Delta_g$ states, a $J$-dependent contribution to the spin-orbit coupling is induced, with a similar $J$ dependence to $A_D$, and this is the most likely the major contributor to $A_D$.

Comparing the results of this first fit with previous fits (tables 1-4) indicates the parameters are essentially consistent. The Hamiltonians used are not identical, though the differences are minor, and mainly relate to differences in definitions of the parameters; see the footnotes to the table for the conversions used for comparison.

It is clear from these plots that all the constants show a smooth trend with v, and an alternative fit was tried, where constants were expressed as polynomials in v + ½, and the polynomial coefficients were fitted rather than constants for individual states. This approach worked well for the smaller constants, such as the spin-spin constant, λ, but required high order polynomials for origin, $B$ and $D$ making them of little value, so this was not pursued, particularly as it is not likely to give a good extrapolation to higher vibrational levels.

Instead, the *ab initio* potential energy curves described above were used as a basis for a global fit. For this the DUO program [37] was used, which allows coefficients in the potential energy expressions above to be fitted to observations. While DUO can fit directly to observed transitions, we chose to fit to rovibrational energy levels calculated by PGOPHER, as it was more convenient to fit the electronic states separately, and we also did not have line lists from some of the earlier work, but could calculate energy levels from published constants. The DUO calculations included much the same rotational terms in the Hamiltonian as the PGOPHER calculations, though the inclusion of the Schrödinger equation for vibrational motion meant that centrifugal distortion parameters were replaced by $r$ dependence of the parameters. The specific parameters included, in addition to the potential, were λ(*r*) and γ(*r*) for all four states, $A(r)$ and $o(r)$ for the B$^3\Pi_g$ and W$^3\Delta_u$ states and $p(r)$ and $q(r)$ for the B$^3\Pi_g$ state. For most terms a linear dependence was assumed:

$$\lambda(r) = \lambda_0 + \lambda_1(r - r_{\text{ref}}) \tag{9}$$

For $A(r)$ the *ab initio* values were taken directly and cubic splines used for interpolation. The input to the fit included calculated levels for all the states measured here, with additional coverage from earlier work: v = 10 – 21 of the B $^3\Pi_g$ state from [14, 19], v = 9 – 16 for the A $^3\Sigma_u^+$ state from [14, 19] and v = 8 – 9 for the B' $^3\Sigma_u^-$ state from [32].

Without adjustment, these calculations gave electronic state separations out by ~100 cm$^{-1}$, vibrational spacings out by ~10 cm$^{-1}$ and rotational energies out by ~10 cm$^{-1}$ at $J$ = 30. The residuals showed smooth trends with v and $J$, and are excellent for what are essentially purely *ab initio* calculations. The residuals could all be reduced to < 2 cm$^{-1}$ by floating just three coefficients in the potentials, specifically $V_{\text{min}}$, $D_e$ and $r_e$, with the exception of the B $^3\Pi_g$ state, where the particularly extensive range of vibrational levels required floating $a_0$. The fitted coefficients are given in table S2. The residuals for all states still showed a systematic trend with $J$ and v, but floating additional parameters in the potential energy curve did not improve the fit. We did not take this approach any further since it ignores some interactions between electronic states, which might be expected to introduce effects of this order of magnitude. Given the analytical form (1) and parameters chosen for floating, the resulting potentials retain the form of the *ab initio* potentials, having simply been subject to a uniform shift and scaling. We can therefore have some confidence that they can be used to predict properties for levels outside the observed range of v and $J$. The adjustment to the B $^3\Pi_g$ potential energy curve is greater than the others, but the observed levels cover 80 % of the well depth.

The fine-structure constants were not varied as part of the fitting process above, as the residuals from varying these parameters were swamped by the systematic residuals from the rotational structure fit. The *ab initio* spin orbit constants, $A$, only required a uniform scaling of 1.003 (for the B $^3\Pi_g$ state) and 1.009 (for the W$^3\Delta_u$ state) for a good fit to the observed values. For the remaining fine-structure terms, $\lambda$, $\gamma$, $o$, $p$ and $q$, the coefficients in equation (9) were determined by manual adjustment so that, when averaged over the vibrational wavefunctions as described below, the observed vibrational dependence of the constants was reproduced.

To generate the final rovibronic parameters, selected constants were then calculated from these refined potential energy curves. This was done using Le Roy's LEVEL program [38]. LEVEL and DUO both solve the one-dimensional vibrational Schrödinger equation but LEVEL was more convenient for this phase; electron spin effects were ignored for these particular calculations. As well as vibrational energies, vibronic transition moments were calculated by integrating the *ab initio* transition moments over the vibrational wavefunctions calculated by LEVEL. Values for rotational and centrifugal distortion constants ($B$, $D$, $H$ …) were derived by using LEVEL to calculate rotational energies (including the centrifugal barrier) for 21 different values of $J$ in the range 0 to 55, and then fitting the energies for a particular vibrational level to the standard polynomial $B_v J(J+1) + D_v J(J+1) + \ldots$. The number of terms was chosen to give a fit with an average error of $< 10^{-4}$ cm$^{-1}$. The values do not exactly match the $B_v$ and $D_v$ values given by LEVEL, which are only based on $J = 0$ wavefunctions, but correspond more closely to usual determinations of these parameters from experimental data, and avoid polynomial truncation problems. A similar approach was used to calculate the other parameters, including centrifugal distortion effects by averaging the fitted $\lambda(r)$ over the vibrational wavefunctions for the same range of $J$ values, and then fitting to a similar polynomial for each v:

$$\lambda_v(J) = \lambda_v + \lambda_{Dv} J(J+1) + \lambda_{Hv} J^2(J+1)^2 + \ldots \qquad (10)$$

In the light of the discussion above concerning $A_D$ and $\gamma$, the $A_v$ values were determined by applying the procedure above to $A(r)$, but only including low $J$ values so that $A_D$ from this source was constrained to zero. $A_D(r)$ was then treated as a separate interaction; in principle this is $r$ dependent, but measurements indicated this was constant to our precision. A similar approach could be used to account for centrifugal distortion of the transition moments, but was not as initial calculations from the *ab initio* curves, and more detailed work by Mishra et al[39], found these effects to be < 2% of the intensity, and thus below our likely error bars.

The parameters generated as above were used as a basis for our final set of recommended parameters, given in tables S5-S8. For practical use, some adjustment is required, given the limits to the potential energy fit. With no adjustment, our observations were reproduced with an overall weighted error of 8200 (0.7 cm$^{-1}$ unweighted). Optimizing just the two dominant constants, the origin and $B$, for all states reduced the overall weighted error to 190, and the unweighted error (0.01 cm$^{-1}$) to less than the linewidth. The largest contribution to the error was now $A$ for the B $^3\Pi_g$ state, and optimising this parameter reduced the error to 11 (0.004 cm$^{-1}$ unweighted). For our recommended values (tables S5-S8) we additionally optimised $\lambda$ and $\gamma$ for v = 0 – 9 of the B $^3\Pi_g$ state, $D$ for v = 0 and 1 of the A$^3\Sigma_u^+$ state and $o$ and $p$ for v = 0 and 1 of the B $^3\Pi_g$ state. (A particularly large range of $J$ values have been measured for v = 0 and 1 for the A and B states.). This set of constants is almost as good as our traditional fit, with all parameters floated, giving a weighted error of 3.8 compared to 2.1 (0.0026 cm$^{-1}$ compared with 0.0023 cm$^{-1}$). Given the minor adjustment to the potential energy curves required, the predictions are likely to have predictive value outside the observed range of v and $J$.

The assignment process above requires a good intensity model, which we now discuss. The Boltzmann plots for individual bands, such as the one illustrated in Fig 3, all show curvature at high $J$, indicative of a non-equilibrium high temperature component. This could be fitted with populations modelled as the sum of two Boltzmann distributions, with two different temperatures. Fitting to a single strong band in each spectrum gave values that worked for all the bands in that spectrum. The values used for each spectrum are given in Table 6.

Table 6. Effective temperatures (K) used in simulating spectra

|  | $T_1$ | $T_2$ | $r$[a] | $T_{vib}$ |
|---|---|---|---|---|
| This Work | 487 | 1161 | 0.029 | 4850 |
| Low Power (25W) spectrum from B&R | 576 | 1384 | 0.062 | 8530 |

a The rotational Boltzmann equation is taken as $\exp(-E/kT_1) + r \exp(-E/kT_2)$

We can also calculate the relative intensities of each band, given the *ab initio* transition moments and adjusted potentials described above. The transition dipole moment for each band is easily calculated by integrating the transition dipole moment function, μ($r$) over the vibrational wavefunctions, and these are tabulated in tables S9-S11. These should predict all the relative intensities for each set of transitions involving a given upper level, and indeed do, within the ±10% spread for transitions within a given band. We found that we could account for the relative intensity between most bands with a separate vibrational temperature, given in Table 6. See Figure 5 for a plot of observed/calculated intensity ratio for the spectrum taken with the lowest power in [7].

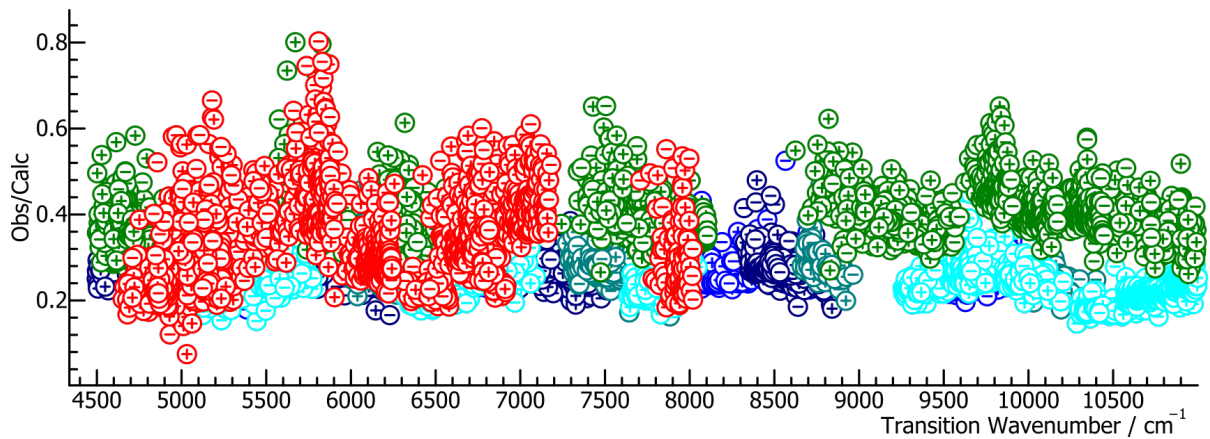

Figure 5. Plot of observed/calculated intensities for the spectrum from [7] taken with the lowest power (25 W). The colours represent different upper electronic and vibrational states; the rotational temperature was modelled with two temperatures (576 and 1384 K) and a vibrational temperature of 8530 K. The W $^3\Delta_u$ – B $^3\Pi_g$ 3–0 band has been excluded from the plot, as the intensity corrections are not available.

The above process gives absolute absorption cross sections, and we can validate these for the B $^3\Pi_g$ – A $^3\Sigma_u^+$ transition by comparing with measured overall lifetimes for vibrational levels of the B $^3\Pi_g$ state. The lifetime can be calculated by summing the Einstein $A$ coefficients for a given upper state over all accessible lower states. (For the B $^3\Pi_g$ state this includes the X $^1\Sigma_g^+$, B' $^3\Sigma_u^-$ and W $^3\Delta_u$ states, but the total rate to these is very small.) There are various measurements of these lifetimes; the most recent of these is by Eyler and Pipkin[40] and, as shown in Table 7, their measured values are entirely consistent with our calculated values.

Table 7 Lifetimes (/μs) for v = 0 – 12 for the B $^3\Pi_g$ state of $N_2$.

| v | This work (Calculated) | Eyler & Pipkin | Error Bar |
|---|---|---|---|
| 0 | 11.94 | | |
| 1 | 9.63 | | |
| 2 | 8.13 | | |
| 3 | 7.08 | | |
| 4 | 6.31 | | |
| 5 | 5.73 | 5.87 | 0.21 |
| 6 | 5.27 | 5.34 | 0.17 |
| 7 | 4.91 | 5.05 | 0.16 |
| 8 | 4.62 | 4.72 | 0.15 |
| 9 | 4.39 | 4.41 | 0.15 |
| 10 | 4.20 | 4.33 | 0.17 |
| 11 | 4.06 | 4.19 | 0.17 |
| 12 | 3.94 | 4.11 | 0.21 |

It is clear from this that the agreement is very good, and indeed for simple small molecules the uncertainty in the theoretical transition moment is likely to be less than the experimental uncertainties in the lifetimes. (See for example reference [41], where ab initio calculations are used as the basis for calculations of absolute transition intensities.) The intensities given in our final line list are thus likely to be accurate to a few %. Absolute intensities have also been calculated on a purely theoretical basis recently by Ni et al [42, 43]; their transition moments are essentially identical to ours, and the calculated results are consistent, though their tabulation of rotational line strengths is less useful as the spin splitting is ignored.

For the other states there are fewer values available for comparison. Some relatively old ab initio calculations are available [44], which are essentially consistent with ours (within 0.02 D) given they are not done to such a high level. Covey et al [45] estimated the strength of the $B\ ^3\Pi_g - W\ ^3\Delta_u$ transition by measuring the Einstein $A$ coefficient of the 2 – 0 band as 0.061 of that of the $B\ ^3\Pi_g - A\ ^3\Sigma_u^+$ 2 – 4 band which is in good agreement with our calculated ratio of 0.081 considering that their estimate of uncertainty of up to ±50%. Lofthus and Kripenje [46] quote an estimate for B' state lifetimes for v = 0 – 8 in the range 25 – 52 μs, consistent with our calculated range of 17 – 47 μs.

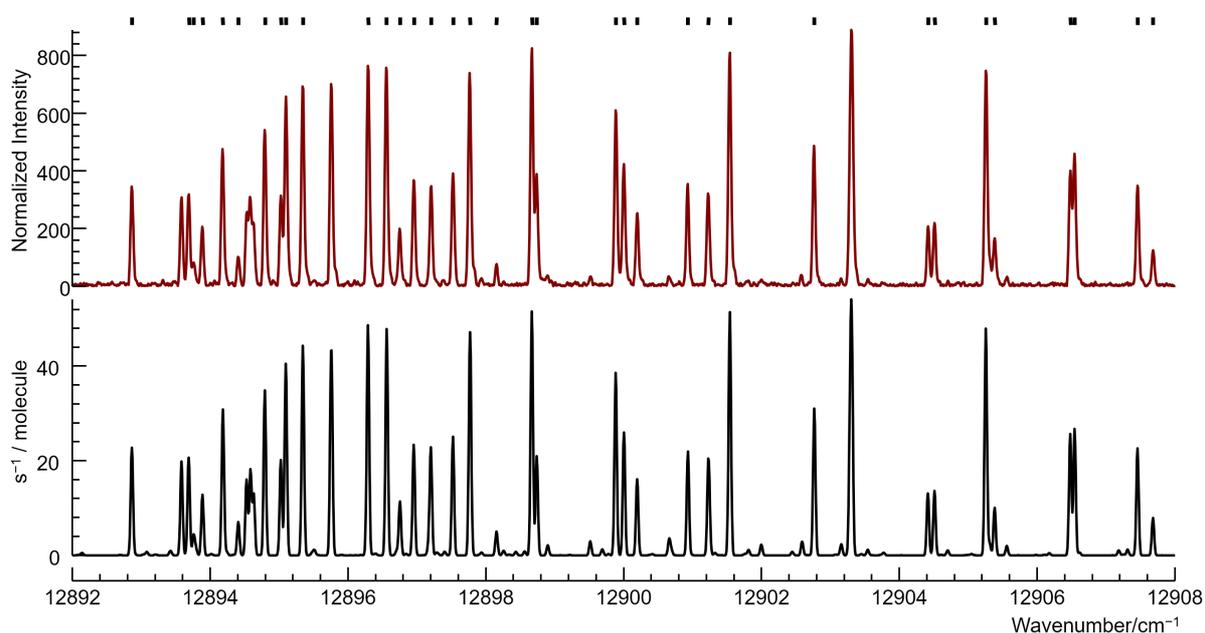

Figure 6. Comparison between part of the FTS experimental spectrum (upper trace) and the PGOPHER simulation (lower trace) in the B-A (2-0) region. The populations were calculated using the parameters given in Table 6. The tick marks at the top of the plot indicate transitions used in the fit; note the excluded transitions are also well modelled. This includes some strong transitions, and all the excluded transitions are blends.

Figure 6 shows the quality of our final fit, comparing a small part of the FTS and simulated spectra. This global fit encompasses all transitions between states A $^3\Sigma_u^+$, B $^3\Pi_g$, B' $^3\Sigma_u^-$ and W $^3\Delta_u$ of the $^{14}N_2$ isotopologue and is based on 17967 observed lines selected to avoid blends using the intensity criterion described above. The derived constants are thus more precise and also more standard than earlier partial contributions concerning B-A [13, 14], B'-B [17], or W-B [18] systems; the new model allows a global description of the main features of the spectrum of $N_2$ between 4500 and 15700 cm$^{-1}$. Experimental line lists and molecular constants are provided as supplementary materials. Figure 7 illustrates the relative importance of the B' $^3\Sigma_u^-$ – B $^3\Pi_g$, W $^3\Delta_u$ – B $^3\Pi_g$ and B $^3\Pi_g$ – A $^3\Sigma_u^+$ systems to the $N_2$ emission spectrum discussed here, plotting emission rates per molecule as calculated for a rotational temperature of 500 K and vibrational temperature 5000 K.

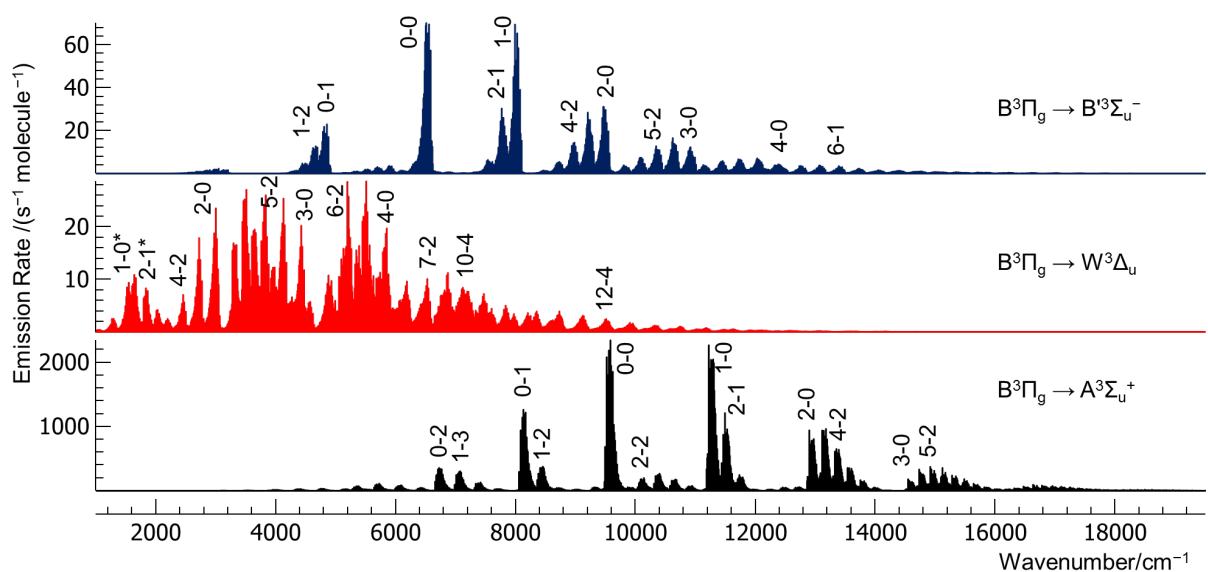

Figure 7. Overview of the contributions of the electronic systems B' $^3\Sigma_u^-$ – B $^3\Pi_g$, W $^3\Delta_u$ – B $^3\Pi_g$ and B $^3\Pi_g$ – A $^3\Sigma_u^+$ to the first positive bands of $N_2$. Note the different vertical scale for each system. Selected vibrational assignments are indicated on the diagram; * indicates emission from W to B rather then B to W.

## 5. Applications

We have two illustrations of practical implementation of this $N_2$ atlas, based on recognition of features in the emission spectrum, without needing line assignments. The first is to calibrate an intracavity absorption spectrum, recorded around the peak of the gain curve of a mode-locked Ti:sapphire laser. The light source is a spectrally broad frequency comb, which is injected into a high finesse optical cavity ($F_c$=17000), whose mirrors curtail the extent of emission emerging from the optical cavity (still of the order of 1000 cm$^{-1}$). The high-reflectivity mirrors produce a large effective pathlength of absorption for any species contained in the cavity, making this a high-sensitivity absorption experiment. An optical Vernier scheme is used to perform continuous optical filtering as the laser frequency comb is scanned across the absorption bands of the sample, analysing the full spectral range at GHz resolution on timescales of the order of 1 second [6]. The optical frequency comb at ILM is not (yet) locked to any external frequency reference, so does not offer metrological precision. Nevertheless, the arrangement combines fast recording times with enough sensitivity to observe absorption from discharge products, and is still an attractive tool for exploratory spectroscopy, except that there is an obvious need to connect the raw time-scale axis of detection (as the Vernier scans) to a reliable wavelength scale. By admitting a small air leak into a discharge fitted within the optical cavity, $N_2$ lines can supply the missing calibration. Figure 8 shows a comparison between a record of $N_2$ B-A bands seen in absorption in this Vernier spectrometer, and the PGOPHER simulation, which has provided the wavenumber scale.

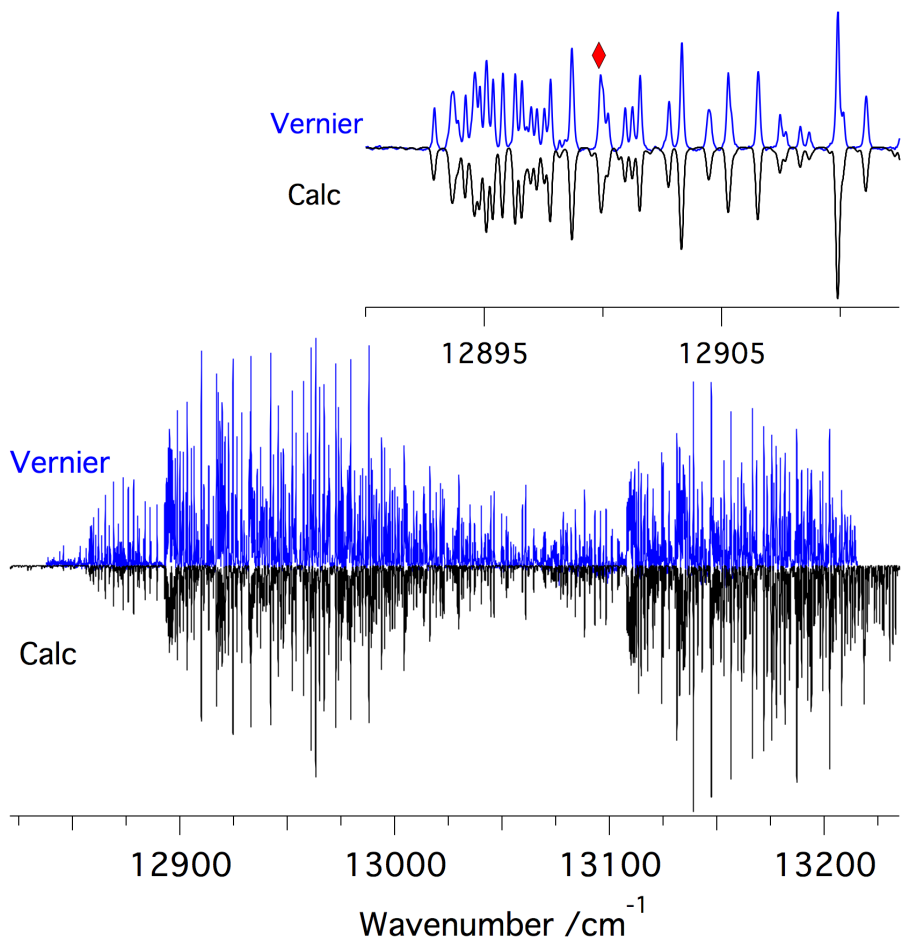

Figure 8. Comparison between Vernier spectrum (upper (blue) trace) and PGOPHER simulation of the $^{14}N_2$ B-A (2-0) band (black trace). The discharge was operating with 0.5 Torr in pure $N_2$, with 50 mA DC current; the simulation was performed setting T = 380 K, and a Gaussian FWHM of 0.18 cm$^{-1}$. The red diamond highlights an overlap of 3 lines (see Fig 7).

The second practical implementation of this $N_2$ atlas is more general, since it offers an alternative to either hot (500 °C) iodine [47] or alkali vapours [48] to calibrate laser excitation spectra in the near IR (our particular concern is with Ti:sapphire-accessible wavelengths, 720 – 1000 nm). Without access to a high-finesse cavity, only weak absorption signals would be expected for transitions from a (metastable) excited electronic state, but optogalvanic spectroscopy is sensitive enough to respond to resonant laser light. This has been demonstrated with cw lasers in the visible region of the B-A system [49, 50], and in the near IR around 775 nm [51]. Feldmann [49] explored optogalvanic signals at modest resolution, tracking changes in the discharge voltage when laser light crossing the discharge tuned through molecular transitions from 570 – 620 nm. He observed changes in sign, and a region of reduced response around 585 nm. Phase effects were also demonstrated at higher resolution in optogalvanic spectra recorded by Bachir *et al.* [51] in a corona-excited supersonic expansion of argon seeded with $N_2$. Figure 2 of their paper showed a 6 cm$^{-1}$ spectral window where lines of the (2-0) band of the B-A system of $N_2$ were of opposite phase to transitions in the (8-7) hot bands. Ref. [51] mentions that the relative phase of optogalvanic signals are sensitive to discharge conditions. To verify that these bands could really be used for secondary calibration purposes, we recorded a few optogalvanic signals around 780 nm at U. New Brunswick, using a home-made hollow cathode glass discharge cell operating at room temperature, and electric circuitry copied from the model developed by Stoicheff and co-workers [52, 53] to detect and amplify the weak variations of the resistivity of the discharge. A slow leak of air flowed through the system, maintaining a pressure ~0.5 Torr around the tungsten anode and stainless-steel cathode. The single-mode cw Ti:sapphire laser beam (output power around 450 mW from a CR 899 cavity) was mechanically modulated at 820 Hz so that signals could be extracted by means of a lock-in amplifier. Clear optogalvanic signals in the *B-A* system of $N_2$ were observed when the pressure was adjusted so that the mauve $N_2$ plasma just filled the hollow cathode. A stable discharge regime was achieved with a current of 15 mA, a DC voltage of about 2000 V and a 115 kΩ ballast resistance. Figure 9 shows a comparison between part of the FTS discharge emission spectrum and optogalvanic under the conditions described above.

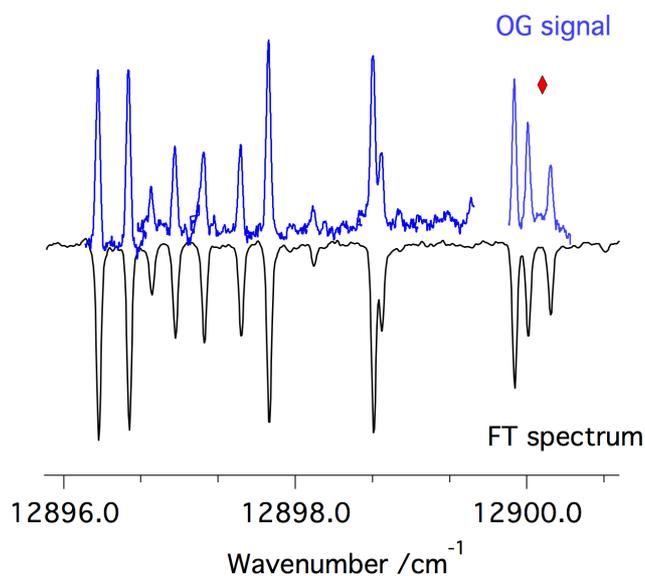

Figure 9. Comparison between part of the FTS discharge emission spectrum (lower trace) and optogalvanic signals (upper trace). The three resolved lines around 12900 cm$^{-1}$ correspond to the single peak also indicated by a (red) diamond in Figure 8.

In the context of high-resolution spectroscopy, where individual lines can be distinguished, an $N_2$ discharge provides a stable and affordable reference spectrum with a high density of lines for calibration purposes. The upper trace in Fig. 9 assembles six individual scans of the Ti:sapphire laser. A single scan (typically < 1 cm$^{-1}$) can easily fall between transitions listed in reference spectra from atomic lamps such as Th/Ar [54] or U [55, 56] for the near IR, as illustrated in Figure 10. Mapping reference signals on to the richer $N_2$ spectrum makes it much easier to establish a reliable wavenumber scale.

We consider that molecular nitrogen could well be considered as a convenient reference, not only in the laboratory, but also for forthcoming high-resolution near IR telescope-based spectrographs, such as SPIRou[57].

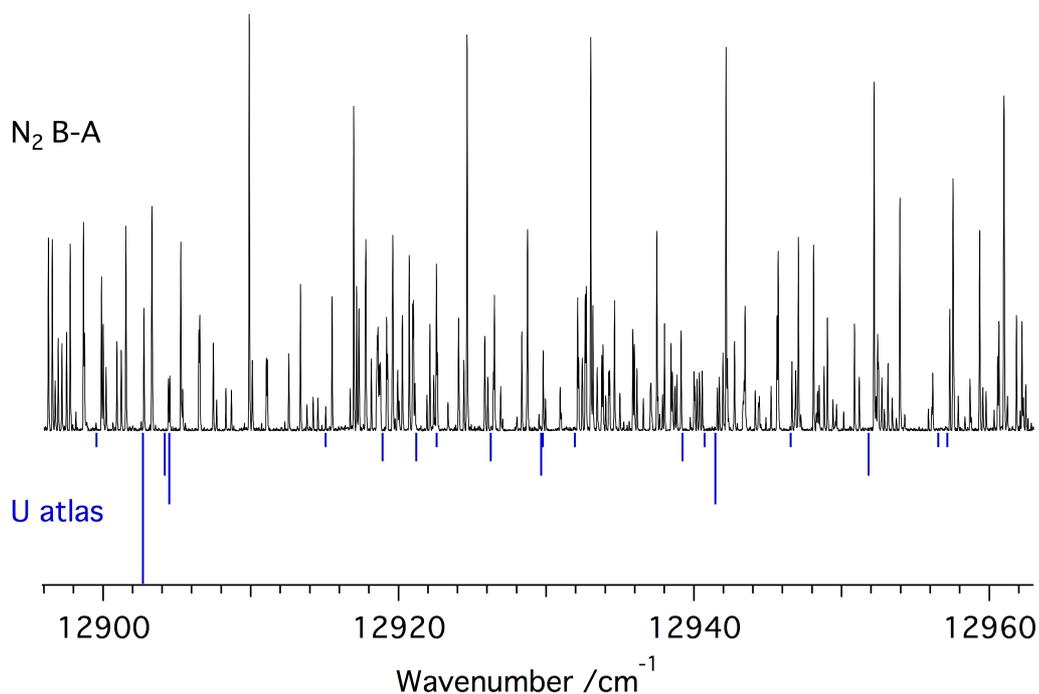

Figure 10. Comparison of a section of the $N_2$ discharge spectrum (upper trace, this work) and U atomic lines from the Uranium atlas [56] (frequently used for calibration of Ti:sapphire laser excitation spectra).

## 5. Conclusion

The $N_2$ bands recorded in [7] have been analysed together with an FT emission spectrum covering 'Ti-sapphire' wavelengths using the PGOPHER program, providing constants for the A $^3\Sigma_u^+$, B $^3\Pi_g$, B' $^3\Sigma_u^-$ and W $^3\Delta_u$ states, involved in the emission spectrum of the first positive system of nitrogen. These are essentially consistent with those reported previously in the 1980's, but more precise, and the Hamiltonian used is more standard. The reference spectrum is available either as an ascii file (9000-15700 cm$^{-1}$), or in parameterized format (PGOPHER). Molecular constants for the *A, B, B'* and *W* electronic states of $N_2$ obtained with PGOPHER describe the spectrum between 4500 and 15700 cm$^{-1}$, identify all observed transitions, and have allowed us to create a reference spectrum for desired conditions (temperature, resolution, lineshape). The global treatment here offers a closer representation of spectral features than earlier analyses. Similar work will be required for taking into account the other spectral features observed in the recorded FTS spectrum, such as those involving $^{15}N_2$ and $^{14}N_2^+$.

**Supplementary Material**:

Selected supplementary material for this paper has been deposited with the Journal, and a full set is available from the University of Bristol Research Data Repository, doi:10.5523/bris.1u7h3szdcpk5r2dfa1hao9o2u1. The files deposited with the Journal include a PGOPHER input file, the calibrated Fourier transform spectrum taken as part of this work, a sample calculated line list for the origin band of the B-X transition and supplementary tables S1 – S13, containing ab initio results and constants not given elsewhere. The additional files in the data repository include a complete line list, *ab initio* output files and input files to the fitting process.

**Acknowledgments**

This work was supported by the French INSU/PNPS (Programme National de Physique Stellaire) program. AJR acknowledges support from the Harrison-McCain foundation as visiting scientist at the University of New Brunswick in 2017. DWT thanks Joyce MacGregor for her assistance with the laser optogalvanic experiments, and the Natural Sciences and Engineering Research Council of Canada for financial support.

**Appendix: Correlation between $A_D$ and $\gamma$**

In Ref. [20], Brown and Watson show that the parameters $\gamma$ and $A_D$ produce indistinguishable contributions to the energy levels of diatomic molecules in $^2\Lambda$ ($\Lambda > 1$) states of diatomic molecules, and therefore only either one or the other of the two parameters can be used in the calculation of energy levels for such states. Their conclusion is subject to some constraints imposed by their approach, in which a zeroth-order Hamiltonian

$$H_0 = T_e + A\hat{L}_z\hat{S}_z + B\hat{\mathbf{N}}^2 \qquad (11)$$

is diagonalized to provide zeroth-order wavefunctions $\left|\psi_n^{(0)}\right\rangle$, and the effects of a perturbing Hamiltonian:(12)are dealt with via first-order perturbation theory. Since this procedure involves taking expectation values $\left\langle\psi_n^{(0)}\left|H_1\right|\psi_n^{(0)}\right\rangle$ over the zeroth-order wavefunctions, some off-diagonal matrix elements of the form $\left\langle\psi_m^{(0)}\left|H_1\right|\psi_n^{(0)}\right\rangle$ for m ≠ n are neglected; these arise from the $N$-dependence of each term in $H_1$. Use of the eigenfunctions of $H_0$ as the basis for further calculations reduces the magnitude of the off-diagonal matrix elements, but since they are not eliminated, data of extensive scope and of high precision may be sensitive to their contributions to the energy when the full Hamiltonian is diagonalized. Under such circumstances the argument for the equivalence of $\gamma$ and $A_{Dv}$ in $^2\Lambda$ states breaks down, and both parameters may be required to fit the data to experimental precision.

In triplet states such as in the $B^3\Pi_g$ state of $N_2$ considered in this paper, Brown and Watson's development should be extended to include the effects of the spin-spin interaction $\lambda$ in $H_1$ (or $H_0$). To do this formally would require significant algebra, but it is straightforward to test for parameter redundancy numerically and this also allows the argument for the breakdown of the perturbation treatment to be quantified. To this end a series of fits were performed on a synthetic energy level list generated with parameters appropriate to $B^3\Pi_g$ v = 0. Λ-doubling effects were ignored, and levels up to $J = 50$ were included. The fits were done for both $^2\Pi$ and $^3\Pi$ states. The "Input" column in the table below gives the parameters used to generate the energy level list; the subsequent columns give the results of fits with $A_D$ or $\gamma$ constrained to zero. As can be seen from the table, a good fit can be obtained for both $^2\Pi$ and $^3\Pi$ with just one of $A_D$ or $\gamma$, but the fit is not perfect, with levels with $J < 10$ showing the largest errors. Note that the origin and $\lambda$ are floated as both $A_D$ and $\gamma$ have some small $J$-independent matrix elements, and fixing the origin or $\lambda$ gives average errors ten times larger.

Table A1 Results of alternative fits to synthetic energy level lists. All values are in cm$^{-1}$.

| | Input | Output for $^2\Pi$ state | | Output for $^3\Pi$ state | |
|---|---|---|---|---|---|
| | | Fix $\gamma$ | Fix $A_D$ | Fix $\gamma$ | Fix $A_D$ |

| Origin | 0 | 2.171(21) × 10$^{-3}$ | 2.471(21) × 10$^{-3}$ | 5.788(29) × 10$^{-3}$ | 6.587(29) × 10$^{-3}$ |
|---|---|---|---|---|---|
| $B$ | 1.6 | 1.600000007(18) | 1.600000007(18) | 1.600000014(24) | 1.600000014(25) |
| $A$ | 40 | 39.995181(60) | 39.994573(60) | 39.995159(51) | 39.994551(51) |
| $\lambda \times 10^3$ | 0 | - | - | −2.252(48) | −2.554(48) |
| $\gamma \times 10^4$ | −40 | 0 | 5.525(12) | 0 | 5.5159(98) |
| $A_D \times 10^4$ | −4 | −0.4810(10) | 0 | -0.48025(85) | 0 |
| Average Error | | 0.00019 | 0.00020 | 0.00032 | 0.00033 |

**References**

[1] Ashrafi M, Lanchester BS, Lummerzheim D, Ivchenko N, Jokiaho O. Modelling of N$_2$ $^1$P emission rates in aurora using various cross sections for excitation. Ann Geophys. 2009;27:2545-53.

[2] Luque A, Gordillo-Vazquez FJ. Modeling and analysis of N$_2$(B $^3\Pi_g$) and N$_2$(C $^3\Pi_u$) vibrational distributions in sprites. Journal of Geophysical Research-Space Physics. 2011;116.

[3] Bucsela E, Morrill J, Heavner M, Siefring C, Berg S, Hampton D, et al. N$_2$(B $^3\Pi_g$) and N$_2^+$(A $^2\Pi_u$) vibrational distributions observed in sprites. Journal of Atmospheric and Solar-Terrestrial Physics. 2003;65:583-90.

[4] Jenniskens P, Laux CO, Schaller EL. Search for the OH (X $^2\Pi$) Meinel band emission in meteors as a tracer of mineral water in comets: Detection of N$_2^+$ (A-X). Astrobiology. 2004;4:109-21.

[5] Gordon IE, Rothman LS, Hill C, Kochanov RV, Tan Y, Bernath PF, et al. The HITRAN2016 molecular spectroscopic database. J Quant Spectrosc Radiat Transf. 2017;203:3-69.

[6] Rutkowski L, Morville J. Continuous Vernier filtering of an optical frequency comb for broadband cavity-enhanced molecular spectroscopy. J Quant Spectrosc Radiat Transf. 2017;187:204-14.

[7] Boesch A, Reiners A. Spectral line lists of a nitrogen gas discharge for wavelength calibration in the range 4500–11 000 cm$^{-1}$. A&A. 2015;582:A43.

[8] Bernath P. Metal Hydrides in Astronomy. AIP Conference Proceedings. 2006;855:143-8.

[9] Brown JM, Carrington A. Rotational Spectroscopy of Diatomic Molecules. Cambridge: Cambridge University Press; 2003.

[10] Brown JM, Merer AJ. Lambda-type doubling parameters for molecules in Π electronic states of triplet and higher multiplicity. J Mol Spectrosc. 1979;74:488-94.

[11] Brown JM, Cheung AS-C, Merer AJ. Λ-type doubling parameters for molecules in Δ electronic states. J Mol Spectrosc. 1987;124:464-75.

[12] Hirota E, Brown JM, Hougen JT, Shida T, Hirota N. Symbols for fine and hyperfine parameters. Pure and Applied Chemistry. 1994;66:571-6.

[13] Effantin C, Amiot C, Verges J. Analysis of the (0-0), (1-0), and (2-0) bands of the B$^3\Pi_g \to$ A$^3\Sigma_u^+$ system of $^{14}$N$_2$ and $^{15}$N$_2$. J Mol Spectrosc. 1979;76:221-65.

[14] Roux F, Michaud F, Verges J. High-resolution Fourier spectrometry of $^{14}$N$_2$ infrared emission spectrum: Extensive analysis of the B$^3\Pi_g$-A$^3\Sigma_u^+$ system. J Mol Spectrosc. 1983;97:253-65.

[15] Roux F, Michaud F. High-resolution Fourier spectrometry of $^{14}$N$_2$ infrared emission spectrum: Extensive analysis of the w$^1\Delta_u \to$ a$^1\Pi_g$ system. J Mol Spectrosc. 1991;149:441-6.

[16]  Effantin C, d'Incan J, Bacis R, Verges J. High resolution Fourier spectrometry of $^{14}N_2$ infrared emission spectrum: Analysis of the (2-0) band of the $W^3\Delta_u \rightarrow B^3\Pi_g$ system. J Mol Spectrosc. 1979;76:204-20.

[17]  Roux F, Cerny D, Verges J. High-resolution Fourier spectroscopy of $^{14}N_2$ and $^{15}N_2$ infrared emission spectrum: Analysis of the $B'^3\Sigma_u^- - B^3\Pi_g$ system. J Mol Spectrosc. 1982;94:302-8.

[18]  Cerny D, Roux F, Effantin C, d'Incan J, Verges J. High-resolution Fourier spectrometry of $^{14}N_2$ and $^{15}N_2$ infrared emission spectrum: Extensive analysis of the $W^3\Delta_u - B^3\Pi_g$ system. J Mol Spectrosc. 1980;81:216-26.

[19]  Roux F, Michaud F. Investigation of the rovibrational levels of the $B^3\Pi_g$ state of $^{14}N_2$ molecule above the dissociation limit $N(^4S)+N(^4S)$ by Fourier transform spectrometry. Canadian Journal of Physics. 1990;68:1257-61.

[20]  Brown JM, Watson JKG. Spin-orbit and spin-rotation coupling in doublet states of diatomic molecules. J Mol Spectrosc. 1977;65:65-74.

[21]  Western CM. PGOPHER: A program for simulating rotational, vibrational and electronic spectra. J Quant Spectrosc Radiat Transf. 2016;186:221-42.

[22]  Western CM, Billinghurst BE. Automatic assignment and fitting of spectra with PGOPHER. Phys Chem Chem Phys. 2017;19:10222-6.

[23]  Western CM. http://pgopher.chm.bris.ac.uk PGOPHER, a Program for Simulating Rotational, Vibrational and Electronic Spectra. Version 9.1 ed: University of Bristol; 2016.

[24]  Vallon R, Ashworth SH, Crozet P, Field RW, Forthomme D, Harker H, et al. Room-Temperature Metal-Hydride Discharge Source, with Observations on NiH and FeH. Journal of Physical Chemistry A. 2009;113:13159-66.

[25]  Kramida AE, Ralchenko Y, Reader J, NIST_ASD_Team. NIST Atomic Spectra Database (ver. 5.1), [Online]. Available: http://physics.nist.gov/asd [2013, October 30]. National Institute of Standards and Technology, Gaithersburg, MD. NIST Atomic Spectra Database (ver 51), [Online] Available: http://physicsnistgov/asd [2013, October 30] National Institute of Standards and Technology, Gaithersburg, MD. 2013.

[26]  Werner H-J, Knowles PJ, Knizia G, Manby FR, Schütz M. Molpro: a general-purpose quantum chemistry program package. Wiley Interdisciplinary Reviews: Computational Molecular Science. 2012;2:242-53.

[27]  Werner H-J, Knowles PJ, Knizia G, Manby FR, Schütz M, Celani P, et al. MOLPRO, version 2010.1, a package of ab initio programs see http://www.molpro.net. 2010. p. MOLPRO, version 2010.1, a package of ab initio programs see http://www.molpro.net.

[28]  Werner HJ, Knowles PJ. A second-order multiconfiguration SCF procedure with optimum convergence. The Journal of Chemical Physics. 1985;82:5053-63.

[29]  Knowles PJ, Werner H-J. An efficient second-order MC SCF method for long configuration expansions. Chemical Physics Letters. 1985;115:259-67.

[30]  Werner HJ, Knowles PJ. An efficient internally contracted multiconfiguration–reference configuration interaction method. The Journal of Chemical Physics. 1988;89:5803-14.

[31]  Knowles PJ, Werner H-J. An efficient method for the evaluation of coupling coefficients in configuration interaction calculations. Chemical Physics Letters. 1988;145:514-22.

[32]  Roux F, Michaud F. Extension of the analysis of the $B'^3\Sigma_u^- \rightarrow B^3\Pi_g$ and $W^3\Delta_u \rightleftharpoons B^3\Pi_g$ systems of the nitrogen molecule by Fourier transform spectrometry. J Mol Spectrosc. 1988;129:119-25.

[33] Le Roy RJ, Huang Y, Jary C. An accurate analytic potential function for ground-state $N_2$ from a direct-potential-fit analysis of spectroscopic data. The Journal of Chemical Physics. 2006;125:164310.

[34] Le Roy RJ, Pashov A. betaFIT: A computer program to fit pointwise potentials to selected analytic functions. J Quant Spectrosc Radiat Transf. 2017;186:210-20.

[35] Forthomme D, McRaven CP, Hall GE, Sears TJ. Hyperfine structures in the v = 1–0 vibrational band of the B-A transition of $N_2$. J Mol Spectrosc. 2012;282:50-5.

[36] Watson JKG. Rounding Errors in the Reporting of Least-Squares Parameters. J Mol Spectrosc. 1977;66:500-2.

[37] Yurchenko SN, Lodi L, Tennyson J, Stolyarov AV. Duo: A general program for calculating spectra of diatomic molecules. Computer Physics Communications. 2016.

[38] Le Roy RJ. LEVEL: A computer program for solving the radial Schrödinger equation for bound and quasibound levels. J Quant Spectrosc Radiat Transf. 2017;186:167-78.

[39] Mishra AP, Narayanan O, Kshirsagar RJ, Bellary VP, Balasubramanian TK. Fourier-transform spectroscopic study of the rotational intensity distribution in the first positive $B^3\Pi_g$–$A^3\Sigma_u^+$ band system of the nitrogen molecule. J Quant Spectrosc Radiat Transf. 2002;72:665-76.

[40] Eyler EE, Pipkin FM. Lifetime measurements of the $B^3\Pi_g$ state of $N_2$ using laser excitation. The Journal of Chemical Physics. 1983;79:3654-9.

[41] Brooke JSA, Bernath PF, Western CM, Sneden C, Afşar M, Li G, et al. Line Strengths of Rovibrational and Rotational Transitions in the Ground State of OH. J Quant Spectrosc Radiat Transf. 2015.

[42] Ni C, Cheng J, Cheng X. Ab initio calculations for the first-positive bands of $N_2$. J Mol Spectrosc. 2017;331:17-22.

[43] Piper LG, Holtzclaw KW, Green BD, Blumberg WAM. Experimental determination of the Einstein coefficients for the $N_2$(B–A) transition. The Journal of Chemical Physics. 1989;90:5337-45.

[44] Werner HJ, Kalcher J, Reinsch EA. Accurate ab initio calculations of radiative transition probabilities between the $A^3\Sigma_u^+$, $B^3\Pi_g$, $W^3\Delta_u$, $B'^3\Sigma_u^-$, and $C^3\Pi_u$ states of $N_2$. The Journal of Chemical Physics. 1984;81:2420-31.

[45] Covey R, Saum KA, Benesch W. Transition probabilities for the $W^3\Delta_u$–$B^3\Pi_g$ system of molecular nitrogen. J Opt Soc Am. 1973;63:592-6.

[46] Lofthus A, Krupenie PH. The spectrum of molecular nitrogen. Journal of Physical and Chemical Reference Data. 1977;6:113-307.

[47] Gerstenkorn S, Vergès J, Chevillard J. Atlas du spectre d'absorption de la molécule d'iode 11000 - 14000 cm-1. Orsay: Laboratoire Aimé Cotton, CNRS II; 1982.

[48] Ross AJ, Bertrand V, Harker H, Crozet P. A Doppler-limited rubidium atlas in ascii format, 9500 – 12 300cm$^{-1}$. J Mol Spectrosc. 2010;264:78-81.

[49] Feldmann D. Opto-galvanic spectroscopy of some molecules in discharges: $NH_2$, $NO_2$, $H_2$ and $N_2$. Optics Communications. 1979;29:67-72.

[50] Sasikumar PR, Harilal SS, Nampoori VPN, Vallabhan CPG. High resolution optogalvanic spectrum of $N_2$-rotational structure of (11, 7) band in the first positive system. Pramana. 1994;42:231-7.

[51] Bachir IH, Huet TR, Destombes JL, Vervloet M. Laser optogalvanic spectroscopy of $N_2$ from the $A^3\Sigma_u^+$ metastable state in a corona-excited supersonic expansion. Chemical Physics Letters. 1997;270:533-7.

[52] Dube P, Kiik MJ, Stoicheff BP. Optogalvanic spectrum of Ar excimers in a DC discharge with supersonic expansion. Chemical Physics Letters. 1995;234:445-9.

[53] Pirali O, Tokaryk DW. Optogalvanic spectroscopy of the $C''^5\Pi_{ui}$–$A'^5\Sigma_g^+$ electronic system of $N_2$. The Journal of Chemical Physics. 2006;125:204308.

[54]  Kerber F, Nave G, Sansonetti CJ. The Spectrum of Th-Ar Hollow Cathode Lamps in the 691-5804 nm region: Establishing Wavelength Standards for the Calibration of Infrared Spectrographs. The Astrophysical Journal Supplement Series. 2008;178:374.

[55]  Stephen LR, James EL, Gillian N, Lawrence WR, Suvrath M. The Infrared Spectrum of Uranium Hollow Cathode Lamps from 850 nm to 4000 nm: Wavenumbers and Line Identifications from Fourier Transform Spectra. The Astrophysical Journal Supplement Series. 2011;195:24.

[56]  Palmer BA, Keller RA, Engleman Jr R. An Atlas of Uranium Emission Intensities in a Hollow Cathode Discharge. Los Alamos informal report. 1980;LA 8251-MS.

[57]  Artigau É, Kouach D, Donati J-F, Doyon R, Delfosse X, Baratchart S, et al. SPIRou: the near-infrared spectropolarimeter/high-precision velocimeter for the Canada-France-Hawaii telescope.  SPIE Astronomical Telescopes + Instrumentation: SPIE; 2014. p. 13.